\documentclass{aa}  
\usepackage{natbib}
\usepackage{dcolumn}
\usepackage{longtable}

\usepackage{color}
\usepackage{ulem}
\usepackage[switch,modulo]{lineno}

\usepackage{graphicx}
\usepackage{txfonts}
\begin{document}

   \title{Near-Ultraviolet Absorption Distribution of Primitive Asteroids from Spectrophotometric Surveys}

   \subtitle{I. Radial Distribution}

   \author{
E. Tatsumi\inst{1,2,3}
          \and
F. Vilas\inst{4}
          \and
J. de Le\'{o}n\inst{1,2}
          \and
M. Popescu \inst{5}
          \and
S. Hasegawa\inst{6}
          \and
T. Hiroi\inst{7}
          \and
F. Tinaut-Ruano \inst{1,2}
         \and
J. Licandro\inst{1,2}
       }             
                    
\offprints{E. Tatsumi, \email{etatsumi@iac.es}}
   \institute{Instituto de Astrof\'{i}sica de Canarias (IAC), University of La Laguna, La Laguna, Tenerife, Spain
    \and
         Department of Astrophysics, University of La Laguna, La Laguna, Tenerife, Spain
         \and
             Department of Earth and Planetary Science, The University of Tokyo, Bunkyo, Tokyo, Japan
            \and
            Planetary Science Institute (PSI), Tucson, AZ, USA
            \and
            Astronomical Institute of Romanian Academy, Bucharest, Romania
            \and
            Institute of Space and Astronautical Science (ISAS), Japan Aerospace Exploration Agency (JAXA), Sagamihara, Kanagawa, Japan
            \and
            Brown University, Providence, RI, USA
           }

   \date{Received xxxx; accepted xxxx}

  \abstract
   {Hydrated minerals, such as phyllosilicates, on asteroids are important for giving constraints on the temperature or compositional distribution of the early Solar System. Previous studies pointed out the possibility that the absorption in the near-ultraviolet (NUV, 0.35 -- 0.5 $\mu$m) wavelength region is a proxy for hydrated minerals in primitive asteroids. However, the radial distribution of the NUV absorption among primitive asteroids was not revisited after the Eight Color Asteroid Survey.}
   {Our objectives were first to evaluate the possibility for using the NUV absorption as diagnostics of hydrated minerals based on the recent datasets of primitive asteroids and hydrated carbonaceous chondrites, and second to investigate the reflectance spectrophotometry of the primitive asteroids in the NUV as functions of heliocentric distance and size. }
   {The NUV and visible reflectance spectrophotometry of more than 9,000 primitive asteroids was investigated using two spectrophotometric surveys, the Eight Color Asteroid Survey (ECAS) and the Sloan Digital Sky Survey (SDSS), which cover wavelengths down to 0.32 $\mu$m and 0.36 $\mu$m, respectively. We classified asteroids from the main asteroid belt, the Cybele and Hilda zones, and Jupiter Trojans based on Tholen’s taxonomy and described the statistical distribution of primitive asteroid types. We also examined the relationship of the NUV, 0.7-$\mu$m, and 2.7-$\mu$m absorptions among primitive asteroids and hydrous carbonaceous chondrites CI and CM. }
   {We found strong correlations between the NUV and the OH-band (2.7 $\mu$m) absorptions for primitive asteroids and hydrated meteorites, suggesting the NUV absorption can be indicative of hydrated silicates. Moreover, there is a great difference in the NUV absorption between the large asteroids (diameter $d > 50$ km) and small asteroids ($d < 10$ km) in the taxonomic distribution. The taxonomic distribution of asteroids differs between the inner main belt and middle-outer main belt. Notably, the C types are dominating large members through the main belt and the F types are dominating small asteroids of the inner main belt. The asteroids beyond the main belt consist mostly of P and D types, although P types are common everywhere in the main belt. The peculiar distribution of F types might indicate a different formation reservoir or displacement process of F types in the early Solar System. The strongest absorptions of the NUV and 0.7-$\mu$m band were observed in G types, which likely comprise CM-like Fe-rich phyllosilicates. On the other hand, according to the recent sample-return from a F-type asteroid (162173) Ryugu, the F types with the OH-band at 2.7 $\mu$m and the shallow NUV absorption could comprise CI-like Mg-rich phyllosilicates.}
   {}


   \maketitle
%

\section{Introduction}

The current state of asteroid distribution resulted from many incidents in the past: the radial transference triggered by gas giants movements, the catastrophic disruptions among asteroids, mixing of asteroids, solar radiation induced radial scattering, and gravitational interferences through resonances of planets. It is currently widely accepted that asteroids were formed in different places from their current locations and subsequently displaced by the migration of gas and ice giants \citep[e.g.,][]{Morbidelli2005,Walsh2011}. This hypothesis of drastic repositioning of asteroids is supported also by the recent laboratory analyses of meteorites: the bimodal distribution of isotopic ratio between non-carbonaceous and carbonaceous chondrites \citep[e.g.,][]{Warren2011, Kruijer2020, Bermingham2020}. The current distribution of asteroids can be important constraints for the models that bridge and depict the evolution from the beginning to the current states of the Solar System. Hydrated minerals on minor bodies are especially important when discussing the origin of volatiles on Earth \citep{Owen&Bar-Nun1995}. Primitive dark minor bodies which inherit many hydrated minerals from the primordial Solar System are the most plausible sources of the water accreted by the Earth later on \citep{Morbidelli2000}. Moreover, if the water on the Earth is delivered from outside, carbonaceous chondritic asteroids are more plausible than comets \citep{Altwegg2015, Marty2017}. This motivated the recent sample return missions, Hayabusa2 by Japanese Aerospace Exploration Agency (JAXA) \citep{Watanabe2019} and OSIRIS-REx by National Aeronautics and Space Administration (NASA) \citep{Lauretta2019}, to the primitive asteroids (162173) Ryugu and (101955) Bennu, respectively. 

Hydrated asteroids mainly are constituted by phyllosilicates which are a product of aqueous alteration on anhydrous silicate rocks in hydrothermal systems. The presence of phyllosilicates constrains the environmental parameters of the early Solar System, such as water-rock ratio and accretion timing of the planetesimals. Asteroids are usually classified by spectral shapes and albedos \citep[e.g.,][]{Tholen1984, BB2002b, DeMeo2009, Mahlke2022}. The first well-established taxonomy was defined using principal component analysis in the near-ultraviolet (NUV) to visible (VIS) range with albedo by \citet{Tholen1984}. He found three major clusters in the principal component space, S, C-X, and D. Those clusters are well defined and might have formed in different regions in the early Solar System. Primitive asteroids often refer to the C-X and D classes. 
However, the origin of C-X and D classes might be different. 

The D-class asteroids are especially dominant in the Jupiter Trojans \citep{DeMeo&Carry2014}. They are hypothesized to be planetesimals formed in trans-Neptunian orbit and captured into co-orbital motion with Jupiter during the time when the giant planets migrated by removing neighbor objects \citep{Morbidelli2005, Levison2009}. Moreover, D-class asteroids are known to exhibit very red spectra compatible to trans-Neptunian objects and cometary nuclei in the visible wavelength region \citep{Jewitt&Luu1990, Luu1994, Emery&Brown2003, Fornasier2004, Campins2006, Licandro2008, Licandro2018, DeMeo&Binzel2008}. 

More recently the silicate spectral features found in Spitzer mid-infrared spectra of the comet nuclei matched those of D-class Trojans \citep{Kelley2017}. Thus, the formation region and materials for C-X and D classes could be fundamentally different. To date, however, still the observations and especially samples from D class are not enough to conclude the fundamental difference between C-X and D classes. We would like to leave this question for the future sample-return and/or in-situ remote-sensing spacecraft missions such as Martian Moons eXplorer \citep{Campagnola2018}. On the other hand, the differences between C and X complexes are not obvious in the principal component space defined by the NUV to VIS asteroid spectra, and could be continuous. X-class asteroids can be categorized into P, M, and E by albedo value. The P-class asteroids have low albedo similar to C-class ones. This suggests possible continuous compositional variation of C and P classes. 

Among the C complex, \citet{Tholen1984} defined F, B, and G sub-classes. The F class is characterized by blue to flat visible spectral end member of the C complex based on the distance from the central cluster of the C complex in the principal component analysis space. Especially, the members of F classes show less drop-off to the near-ultraviolet wavelength. The B class shows blue visible spectral slope and drop-off into the infrared. Also it is known that the albedo of the B class, such as (2) Pallas, is marginally higher than the typical C class. The G class is assinged to the objects with exceptionally deep ultraviolet absorption. These sub-classes are not well separated from the C complex, and should be considered as end members of the C complex.

The NUV absorption is observed in spectra of both hydrous meteorites and primitive asteroids. The steep absorption shortward of 0.4 $\mu$m is attributable to an Fe$^{2+}$-Fe$^{3+}$ charge transfer band \citep{Gaffey&McCord1979}. Hydrated layer-lattice silicate or clay mineral grains comprise the bulk of CI-CM assemblages. Hydrated mineral grains contain both Fe$^{2+}$ and Fe$^{3+}$ giving rise to an intense charge transfer absorption in the blue, which is evident even with very low albedo asteroids. Other iron-rich silicates such as the olivine and pyroxene in CV and CO meteorites also produce NUV absorption. However, these minerals have lower optical densities than hydrated silicates, which means their UV absorption will be more effectively suppressed by the presence of opaque minerals, and high optical density hydrated silicates can dominate the NUV spectral reflectance \citep{Feierberg1981}. The strong correlation between the NUV and 3 $\mu$m was first suggested by \citet{Feierberg1985} measuring the reflectance at 2.92 $\mu$m instead of the OH-band itself. Most hydrous meteorites, CI, CM, and CR, exhibit a strong absorption in the NUV \citep{Cloutis2011a, Cloutis2011b, Cloutis2012, Hiroi2021}. \citet{Hiroi1996a, Hiroi1996b} showed the correlation between NUV and OH-band for hydrous meteorites, including the heated Murchison (CM2) and Ivuna (CI1) meteorites. More specifically, once meteorites are heated and dehydrated, both the NUV and OH-band absorptions are weakened. They also pointed out that some naturally thermally metamorphosed CI/CM meteorites (ATCC) exhibit much weaker NUV absorptions.

Even though the laboratory measurements suggest the possibility of hydrated silicate measurements by the NUV absorption, the quantitative NUV absorption distribution among asteroids has not been discussed. This might be because of the difficulty in NUV reflectance observations. One reason is the low sensitivity in the NUV for CCDs and rapid decrease of solar photon flux at shorter wavelengths. Moreover, because the Rayleigh scattering by the atmosphere affects more on a shorter wavelength, the observable photon flux will be much smaller and the signal-to-noise ratio (SNR) will be lower for the NUV region from the ground. Another reason is the rareness of well-characterized solar analogs. Spectroscopic measurements of asteroids’ reflectance usually use a solar analog flux observed in a similar sky condition to divide an asteroid flux. However, there are very few well-characterized solar analogs in the NUV \citep{Hardorp1978, Tedesco1982, Tatsumi2022}. Thus, for quantitative spectroscopic measurement in the NUV, the solar analogs need to be investigated first. There are a few attempts to do this with ground-based spectroscopic studies in the NUV \citep{Tatsumi2022}, however, the possible uncertainties or errors have not been adequately understood. On the other hand, the photometric studies are more reliable in the NUV, because the photometric filters are well studied and characterized by numerous standard stars as well as the high SNR due to the broad bands. Thus, the well-defined asteroid reflectance colors in the NUV photometric surveys are suitable datasets in our investigation. In this study, we investigate the NUV distribution among asteroids in the main asteroid belt to the Cybele and Hilda regions, and discuss the distribution of hydrated asteroids. Finally, we will discuss the formation of primitive asteroids/planetesimals. Our study can be a milestone for the future NUV investigation of asteroid surveys such as the Gaia DR3 spectroscopic data \citep{Galluccio2022,Tanga2022} and the J-PLUS data \citep{Morate2021}.

\section{Diagnostics of hydrated minerals in other wavelength regions}
Hydrated minerals among asteroids have been investigated by reflectance spectroscopy. Direct indication of hydrated minerals can be found around the 3-$\mu$m wavelength range. \citet{Lebofsky1978} first detected the 3-$\mu$m absorption on (1) Ceres by ground-based observations, which had been later confirmed by the in-situ observations of the Dawn spacecraft \citep{DeSanctis2015, Ammannito2016}. Later, more primitive asteroids were found to have the 3-$\mu$m absorption (in fact, the large majority of the C and P classes observed in this region have it). There is variation in the depth, center and shape of this band \citep{Lebofsky1980, Lebofsky1990, Feierberg1985, Jones1990, Rivkin1995, Rivkin2002, Rivkin2015, Rivkin2019, Rivkin2022, Takir&Emery2012, Takir2015}. The 3-$\mu$m band is a broad and complex absorption of metal-OH, interlayer water, water ice, NH$_3$-bearing phases, carbonates, and organics. Metal-bearing phyllosilicates, such as serpentine and saponite, exhibit a sharp absorption centered at 2.7 -- 2.8 $\mu$m, here after the OH-band for convenience. The peak wavelength can evolve from the weakly altered CMs (maximum depth at $\sim$2.8 $\mu$m) to the extensively altered CMs/CIs (maximum depth at $\sim$2.7 $\mu$m), attributed to variations in the chemistry of the phyllosilicate phases from Fe-rich to Mg-rich \citep{Beck2010}. The OH-band overlaps with the atmospheric water band and cannot be directly observed from ground-based telescopes. The AKARI space infrared telescope directly investigated this region for 66 asteroids \citep{Usui2019} and it found that 17 out of 22 (77\%) C-complex asteroids and 5 out of 8 (63\%) low-albedo ($p_V<0.11$) X-complex asteroids have significant OH-band absorption \citep{Usui2019}. Only two out of 17 S-complex asteroids show a possible OH-band absorption. One T-class asteroid, (308) Polyxo, shows the OH-band absorption while none of the three D-class asteroids do (although only one T-class was observed by AKARI). It should be noted that, so far, a possible OH-band was detected on only one D-type asteroid, (773) Irmintraud by ground-based observations\citep{Kanno2003}. However, (773) Irmintraud was later observed by AKARI and the OH-band was not confirmed \citep{Usui2019}.

The shallow absorption feature around 0.7 $\mu$m is also known to be associated with Fe-bearing phyllosilicates, which is caused by Fe$^{2+}$-Fe$^{3+}$ intervalence charge transfer. Given the abundance of Fe-bearing phyllosilicates in CMs, many CMs exhibit this feature. The 0.7-$\mu$m band was observed on asteroids as well \citep{Vilas&Gaffey1989, Vilas1993, Vilas1994, Barucci1998}. The strong correlation between the 0.7-$\mu$m band and the OH-band was pointed out by various studies \citep{Vilas1994, Howell2011, Rivkin2015}. AKARI has confirmed that the 0.7-$\mu$m feature always associates with the OH-band \citep{Usui2019}. However, absence of the 0.7-$\mu$m feature does not necessarily mean anhydrous objects \citep{Rivkin2015, Usui2019}. Thus, the presence of the 0.7-$\mu$m band on asteroids is more likely to be a proxy of CM-like mineral composition but not a proxy of phyllosilicates in general. The 0.7-$\mu$m band central wavelength of asteroids are shorter than those of CM meteorites. \citet{Fornasier2014} suggested CM2 meteorites could be only a subset of those asteroids with 0.7-$\mu$m band based on the comparison of central wavelength between CM2 meteorites and primitive asteroids. Alternatively, based on laboratory reflectance spectra, \citet{Vilas1994} proposed that the difference in the central wavelength of the 0.7-$\mu$m absorption feature observed between CM2 meteorites and primitive asteroids is a function of temperature on the asteroid surfaces. It is also possible that the 0.7-$\mu$m band is shallower compared with the depth of the OH-band, meaning that there could be observational bias due to the low signal-to-noise ratio for the 0.7-$\mu$m band. The 0.7-$\mu$m band, however, is modeled to be detectable in a visible spectrum having a signal-to-noise ratio of 10 \citep{Vilas1997}.


\section{Asteroid spectrophotometry dataset}
\begin{figure}[ht]
\centering
\includegraphics[width=\hsize]{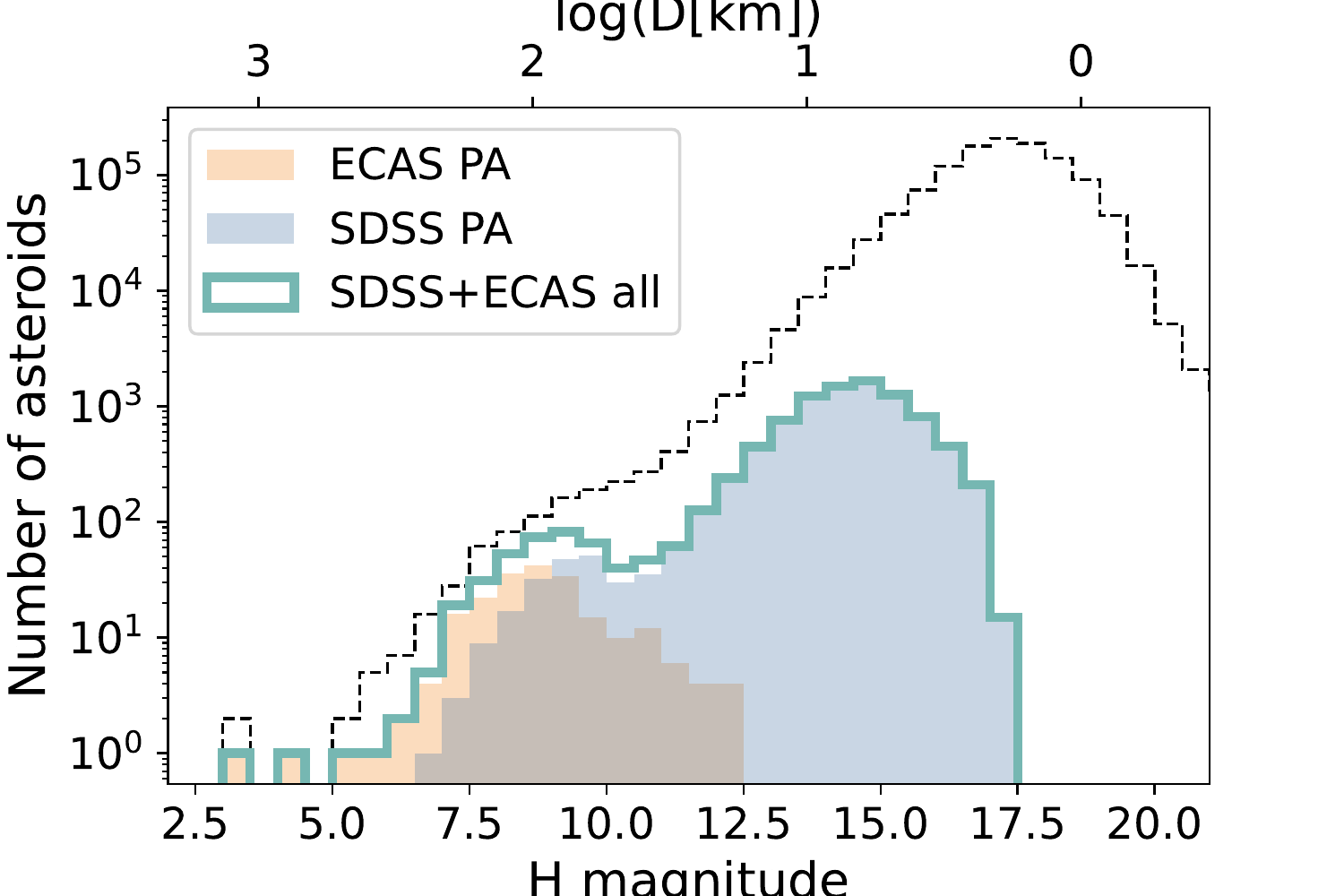}
\caption{Absolute magnitude distribution of asteroids. The distributions of primitive asteroids (PAs) from SDSS and ECAS datasets used in this study are indicated by orange and blue hatches. The integrated PA dataset of SDSS and ECAS is shown by the green line. The black dashed line is the distribution of all asteroids in the ASTORB database (accessed on Dec. 2022). The x-axis on the top corresponds to approximated diameter assuming an albedo value $p_V=0.06$ using Eq. \ref{eq5}, typical of C types.}
\label{fig:sample}
\end{figure}

The number of asteroid observational datasets down to the NUV wavelength range are quite limited so far. Three photometric surveys, Sloan Digital Sky Survey (SDSS), Eight Color Asteroid Survey (ECAS), and J-PLUS had been conducted down to wavelengths $<0.4$ $\mu$m. 
We did not include the J-PLUS data in our analysis because most of the objects are included in the SDSS and ECAS. ECAS obtained photometric information using eight filters: \texttt{s} (0.337 $\mu$m), \texttt{u} (0.359 $\mu$m), \texttt{b} (0.437 $\mu$m), \texttt{v} (0.550 $\mu$m), \texttt{w} (0.701 $\mu$m), \texttt{x} (0.853 $\mu$m), \texttt{p} (0.948 $\mu$m), and \texttt{z} (1.041 $\mu$m) \citep{Zellner1985}, and SDSS using 5 filters: $u$ (0.3557 $\mu$m), $g$ (0.4825 $\mu$m), $r$ (0.6261 $\mu$m), and $i$ (0.7672 $\mu$m), $z$ (0.9097 $\mu$m) \citep{Fukugita1996}. The SDSS and ECAS surveys are complementary in terms of the absolute magnitude of the objects, meaning that SDSS mostly observed $8<\mathcal{H}<20$ and the objects in ECAS are mostly $\mathcal{H}<10$ (Fig. \ref{fig:sample}). This is because SDSS has the unique attribute that it is not able to measure the largest, brightest asteroids. Thus, we used these two datasets to cover a wide size range of asteroids. In addition, we also used the NIR spectra obtained by the AKARI space telescope, and the NIR spectrophotometry obtained by MOVIS.

\subsection{ECAS}
The Eight Color Asteroid Survey (ECAS)\footnote{\url{https://sbn.psi.edu/pds/resource/ecas.html}} is the photometric survey of 589 asteroids, among which 405 astreoids were chosen as high-quality data. They derived color indices of asteroids to give mean color indices of zero for four well-characterized solar analogs \citep{Tedesco1982}. Thus, spectral reflectance $R_\lambda$ can be obtained by $\log(R_\lambda)=\pm 0.4 c_\lambda$, where $c_\lambda$ is the tabulated color index and the negative sign is chosen for wavelengths shorter than \texttt{v} band.

\citet{Tholen1984} developed the taxonomic classification based on cluster analysis in the principal component space applied to the ECAS dataset with known geometric albedo which were given mainly by Tuscon Revised Index of Asteroid Data \citep[TRIAD;][]{Morrison&Zellner1979}.
We used the albedo $p_V$, absolute magnitude $\mathcal{H}$, and diameter $d$ values adapted by the AcuA dataset by AKARI \citep{Usui2011} and the NEOWISE dataset \citep{MainzerPDS}, which are the newer datasets and cover more numbers of asteroids, while the original Tholen's taxonomy used the albedo from TRIAD \citep{Bender1978}. The comparison between the AKARI, IRAS and NEOWISE surveys suggests that NEOWISE might overestimate the albedo for large asteroids possibly owing to the detector saturation \citep{Usui2014}. While AKARI completed the albedo survey for the asteroids $\mathcal{H}<9$ and WISE measured many small asteroids which peak at $\mathcal{H}\sim 15$. By updating albedo using these two albedo catalogs, we could exploit the albedo of 536 asteroids out of 589 asteroids in ECAS. Furthermore, we found that there is a significant discrepancy in the albedo values between AKARI+NEOWISE and TRIAD. 

\begin{figure}
\centering
\includegraphics[width=\hsize]{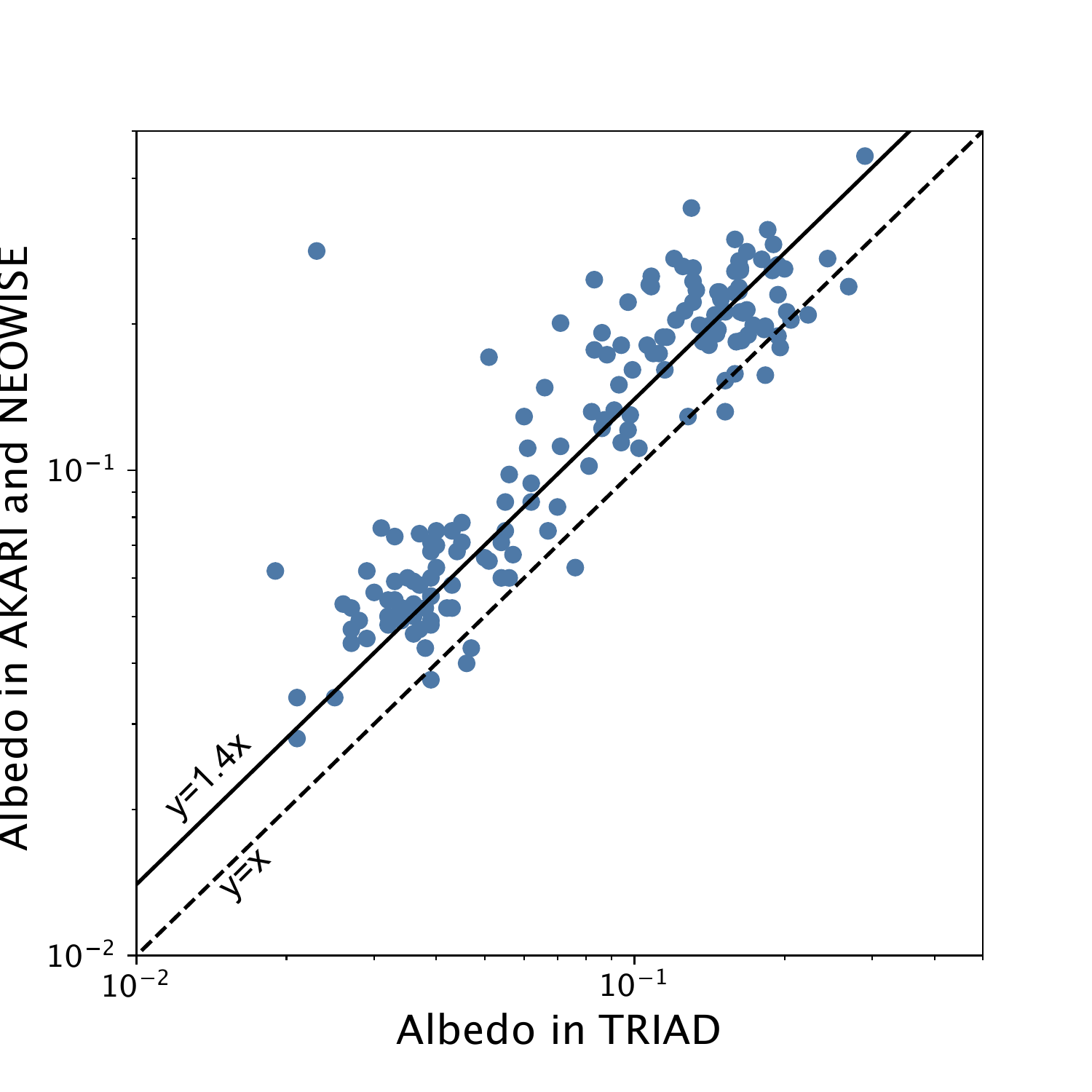}
\caption{Albedo values in TRIAD (data from \citealt{Morrison&ZellnerPDS}) compared with the AKARI+NEOWISE dataset. The solid line indicates a linear fit $y=1.4x$ and the dashed line indicates $y=x$.}
\label{fig:TRIAD}
\end{figure}

Figure \ref{fig:TRIAD} shows the albedo of the same asteroids in AKARI+NEOWISE and TRIAD, suggesting TRIAD dataset may underestimate albedo. Linear fitting of this dataset indicates that the AKARI+NEOWISE values are 1.4 times of the TRIAD values. This affects the classification of E (high albedo), M (medium albedo), and P (low albedo) classes in the X complex using albedo. Although originally the threshold of P class was $p_V<0.08$, using the AKARI+NEOWISE dataset the threshold is better set at $p_V<0.11$. Moreover, the bimodal histogram distribution of X class exhibits the minima around $p_V\sim0.11$ \citep{Usui2013}. Thus, we use the $p_V<0.11$ for classifying P class in this study.

We used high-quality 212 asteroids classified into C-complex, P (low-albedo X), and D classes based on Tholen’s taxonomy with albedo values adopted from AKARI+NEOWISE. We calculated the reflectance at the wavelengths of SDSS by the interpolation of two reflectance values at the nearest wavelengths from ECAS, and then, the spectral slopes were calculated by the following equations:
\begin{align}
    S_{\rm NUV}&=\frac{(R_g/R_u)-1}{\lambda_{g}-\lambda_{u}}, \label{eq1}\\
    S_{\rm VIS}&=\frac{(R_i/R_g)-1}{\lambda_{i}-\lambda_{g}},
\end{align}
where $R$ is the reflectance at the SDSS filter interpolated from two nearest filters of the ECAS filter system, and $\lambda$ is the effective filter wavelength of the SDSS filter. 
In addition to this, we defined the NUV absorption strength as $S_{\rm NUV}-S_{\rm VIS}$.
We also measured the indication of a 0.7-$\mu$m band absorption in the following two ways: \begin{align} 
{\rm HYD_{\rm ECAS}}&=1-\frac{2R_{\tt w}}{R_{\tt v}+R_{\tt x}}, \label{eqhyd}\\
{\rm HYD_{\rm SDSS}}&=1-\frac{2R_i}{R_r+R_z}.
\end{align}
The positive values of these parameters may indicate the stronger absorption at the 0.7-$\mu$m.
The albedo, spectral slope, NUV absorption, and HYD for each taxonomic class are summarized in Table \ref{table:tax}. Some asteroids overlap in multiple classes, thus the total number of samples in all taxonomic classes exceed 212. This result suggests that the VIS spectral slope $S_{\rm VIS}$ and NUV absorption strength can classify those primitive asteroid classes. The G class exhibits the deepest NUV absorption by definition. The C and B classes have similar NUV absorption strengths but they are different in VIS spectral slopes. There are also large differences in albedo, and G and B classes are especially bright among primitive asteroids. Moreover, although the wavelengths of the SDSS filter system are not optimized for measuring the 0.7-$\mu$m band depth, the good correlation between HYD$_{\rm ECAS}$ and HYD$_{\rm SDSS}$ is shown in Table \ref{table:tax}. This agrees with the previous work by \citet{Rivkin2012} showing that the Ch and Cgh asteroids (in SMASS taxonomy), which exhibit the 0.7-$\mu$m band absorption, have positive values of HYD. This suggests that HYD$_{\rm SDSS}$ could be a good indicative for the 0.7-$\mu$m band. Nevertheless, the high HYD value for D types might be because of their concave visible reflectance spectral shapes.
\begin{table*}
\caption{Spectral slopes and albedo for each of Tholen’s primitive taxonomy classes. Note that some asteroids were classified into multiple classes in \citet{Tholen1984} and thus the total number (244) of samples exceeds 212.}
\label{table:tax}
\centering
\begin{tabular}{c c c cc c cc}
\hline\hline
Class & Sample No. & Albedo & $S_{\rm VIS}$& $S_{\rm NUV}$ & $S_{\rm NUV}- S_{\rm VIS}$ & HYD$_{\rm ECAS}$ & HYD$_{\rm SDSS}$\\
 & & (\%) &  ($\mu$m$^{-1}$) &  ($\mu$m$^{-1}$) &  ($\mu$m$^{-1}$) & (\%) & (\%)\\ \hline
 C & 127& $5.9\pm 2.3$ & $0.09\pm0.13$ & $0.90\pm 0.32$ & $0.81\pm0.31$ & $0.3\pm2.4$ & $-0.1\pm1.2$\\
 F & 25 & $5.9 \pm 2.2$ & $-0.04\pm 0.14$ & $0.28 \pm 0.25$ & $0.33\pm0.26$ & $-1.1\pm2.7$ & $-0.7\pm1.1$\\
 G & 8 & $8.0\pm 0.9$ & $0.08 \pm 0.0.07$ & $1.58\pm 0.14$ & $1.58\pm0.14$ & $1.7\pm2.1$ & $0.7\pm0.8$\\
 B & 11 & $9.9\pm3.7$ & $-0.07\pm0.11$ & $0.74\pm0.27$ & $0.81\pm0.33$ &$-1.1\pm1.7$ &$-0.03\pm1.0$\\
P & 52 & $4.9\pm1.4$ & $0.29\pm0.15$ & $0.56\pm0.29$ & $0.31\pm0.31$ &$-1.2\pm1.7$ &$-0.9\pm1.2$\\
D & 21 & $5.1\pm1.5$& $0.91\pm0.11$& $0.49\pm0.27$ & $-0.42\pm0.27$ &$-0.8\pm1.7$ &$-0.4\pm1.1$\\
\hline
\end{tabular}
\end{table*}

\subsection{SDSS}
\begin{figure}
\centering
\includegraphics[width=0.9\hsize]{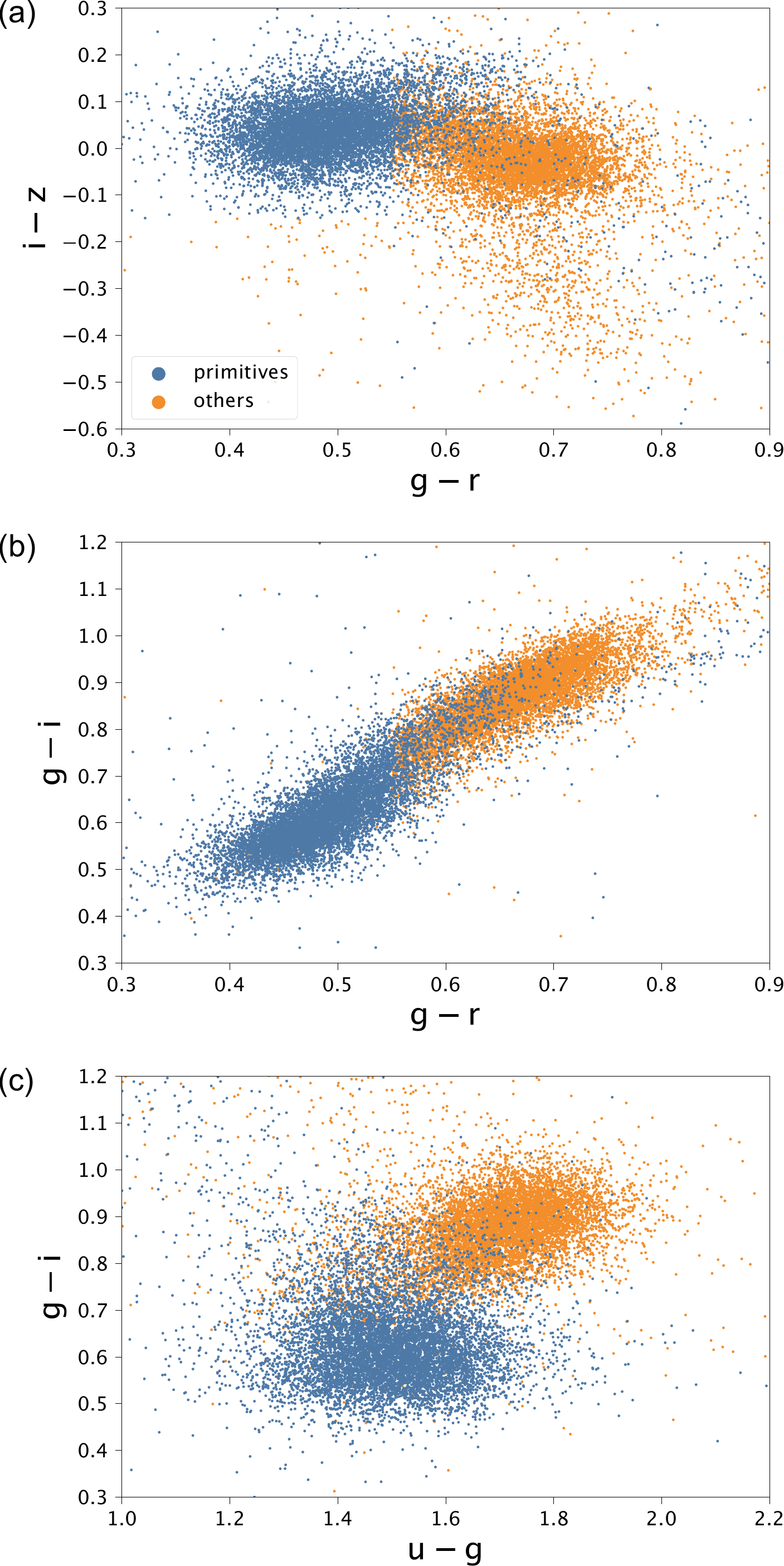}
\caption{Primitive asteroids (blue) separated from other classes (orange) based on our criteria in the color plots: (a) ($g-r$) vs. ($i-z$); (b) ($g-r$) vs. ($g-i$); and (c) ($u-g$) vs. ($g-i$).}
\label{fig:sdss_color}
\end{figure}

The SDSS is the un-targeted broadband photometry survey. There have been several exploitation of solar system objects from SDSS so far.
The SDSS Moving Object Catalog (MOC) is a series of photometric surveys of moving targets, mainly asteroids \citep{Ivezic2002}. The 4th data of SDSS MOC was published in 2008 \citep{Ivezic2010}. This contains more than 100,000 known Solar System objects at that time, although it contains only a fraction of the entire SDSS dataset. More recently, \citet{Sergeyev&Carry2021} conducted an exhaustive search on moving objects in SDSS images including all the observational period. This catalog includes $\sim 380,000$ known Solar System objects and its completeness is estimated to be about 95\%.
We limited the semi-major axis from 2 to 5.2 au to obtain the main belt to the Cybele, Hilda, and Jupiter Trojan zone asteroids. In the same was as for the ECAS dataset, the albedo values were adopted from the AKARI and NEOWISE data. In our analysis we did not include the asteroids without albedo values. 
First, good-quality data was selected based on the photometry flag and the errors in observations. \citet{Sergeyev&Carry2021} defined the photometry flag to discriminate bad data. Moreover, since the $u$-band observations tend to have larger errors, we used objects with errors smaller than 0.1 mag. 
Second, to separate primitive asteroids from the whole dataset, we applied the following thresholds based on the color boundaries of B, C, X, and D complexes calculated by \citet{Sergeyev&Carry2021}: the bluer group ($g-r) < 0.55$ and ($i-z) >-0.15$ for the C complex, and the redder group ($g-r$) $\geq$ 0.55 and $p_V < 0.11$ for P and D classes. This is because it is known that sometimes the B class contains high albedo members \citep[e.g.,]{Tholen1984, Usui2013, Ali-Lagoa2013}. The primitive asteroids selected from the whole dataset are shown in Fig. \ref{fig:sdss_color}.  
Also, we did not use faint samples with absolute magnitude $\mathcal{H} < 17.5$ for main belt asteroids to avoid observational bias in which small, dim objects at a far distance are less observable. 
These selections gave us the final  photometric spectra of 8,956 objects. 
Like for the ECAS dataset, the VIS and NUV spectral slopes were calculated using the following equations:
\begin{align}
    S_{\rm NUV}&=\frac{10^{-0.4(g-u-(g-u)_\odot)}-1}{\lambda_{g}-\lambda_{u}},\\
    S_{\rm VIS}&=\frac{10^{-0.4(i-g-(i-g)_\odot)}-1}{\lambda_{i}-\lambda_{g}},
\end{align}
where $\lambda$ is the effective filter wavelength of the filter and the solar colors are $(i-g)_\odot=-0.57$ , $(g-u)_\odot=-1.40$ from \citet{Holmberg2006}. 

\subsection{Comparison between SDSS and ECAS}
Using the SDSS and ECAS dataset, we obtained the spectrophotometry of 9,168 objects in total. We estimated the diameter from the absolute magnitude. The transformation between the asteroid absolute magnitude $\mathcal{H}$ and its effective diameter $d$ [km] requires knowledge of the albedo $p_V$ \citep{Harris1997},
\begin{equation}
    \mathcal{H}=18.1-2.5\log \left(\frac{p_V}{0.1}\right)-5\log(d).\label{eq5}
\end{equation}

\begin{figure}[ht]
\centering
\includegraphics[width=0.9\hsize]{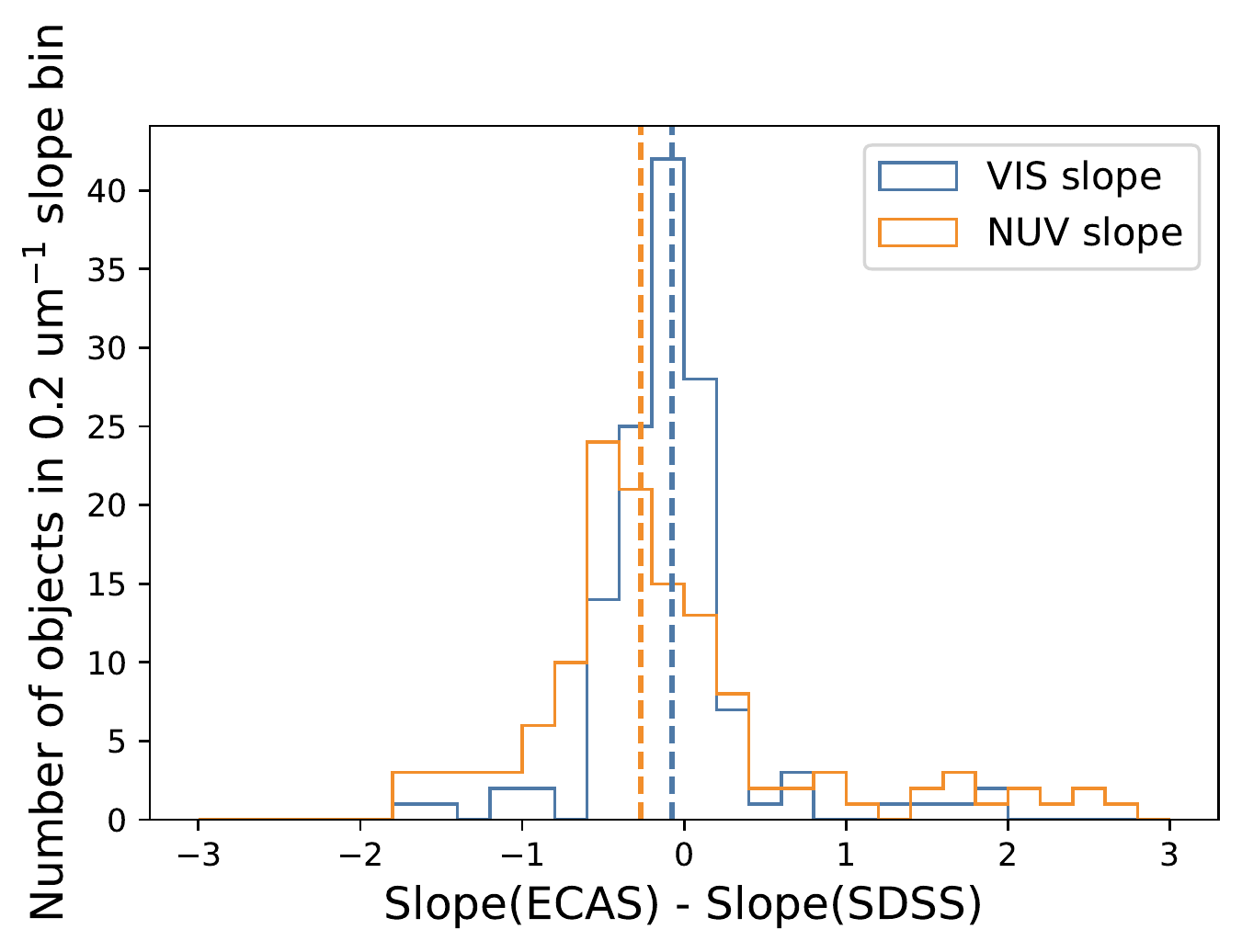}
\caption{Difference of spectral slope values of the common objects between ECAS and SDSS. We measured 132 common asteroids between two catalogs and compared the values. We see a shift of $-0.08$ [$\mu$m$^{-1}$] for VIS slope and $-0.27$ [$\mu$m$^{-1}$] for NUV slope in the SDSS data.}
\label{fig:caldiff}
\end{figure}

Among the asteroids which have been observed multiple times with SDSS, there are 132 objects observed commonly in both SDSS and ECAS (including all taxonomy classes). Thus, we compare the VIS and NUV spectral slopes in both catalogs. We found slight offsets between values from the two catalogs (Fig. \ref{fig:caldiff}). This might be because of uncertainty in the solar color in the broad-band filters, and the interpolation of spectra to match wavelengths of ECAS to SDSS. Thus, we correct this offset by adding the median values of differences in spectral slopes between SDSS and ECAS, $-0.27$ $\mu$m$^{-1}$ and $-0.08$ $\mu$m$^{-1}$ from the NUV and VIS slope values calculated from SDSS, respectively.

\section{Relation between the OH-band and the NUV absorption}

In this section, we tested the relationship between the OH-band and the NUV absorption, which was raised by \citet{Feierberg1985, Hiroi1993, Hiroi1996a, Hiroi1996b}. We used recent datasets from meteorites and asteroids to check if the NUV absorption can be the proxy of hydrated minerals among the primitive asteroids.

\subsection{Meteorites}
Primitive asteroids are linked to carbonaceous chondrites. Especially, the hydration of meteorites and asteroids is thought to occur through aqueous alteration of precursor anhydrous minerals. CM, CI, and CR meteorites are reported to have phyllosilicate abundances of 70 -- 90 vol\%, 81 -- 84 vol\%, and 1 -- 70 vol\%, respectively \citep{Bland2004, Howard2011, Howard2015, King2015}. Some hydrous carbonaceous chondrites experienced subsequent thermal metamorphism \citep[ATCCs;][]{Ikeda1992, Nakamura2005, Tonui2014}. Phyllosilicate abundance of ATCCs varies from 0 to >80 vol\% depending on the degree of heating \citep{King2021}. ATCCs in heating stage IV (heating temperature of >750 $^\circ$C) do not contain phyllosilicates due to decomposition, while ones in heating stage I and II (300 -- 500 $^\circ$C) show similar phyllosilicate abundance as unheated CM/CIs \citep{King2021}. 

In this study we focus on the hydrous carbonaceous chondrites and ATCCs to illustrate the spectral characteristics related to hydration and dehydration states. \citet{Hiroi2021} recently measured the reflectance spectra of 148 carbonaceous chondrites selected from the Antarctic meteorite collections of National Institute of Polar Research in Japan, NASA Johnson Space Center, and Smithsonian Institute National Museum of Natural History. Together with the previous data from the RELAB database (see Table 3 in \citealt{Hiroi2021}), 78 spot spectra from 77 hydrous meteorites (CI, CM, CR, Tagish Lake, and ATCCs) were obtained covering the wavelength of 0.3 -- 4 $\mu$m under ambient air. They conducted 6th-order Gaussian fitting to evaluate the atmospheric water contamination. Water absorption is a broad and round-shaped absorption typically around 3.1 $\mu$m. Here we use the Gaussian fitting bands having a center wavelength shorter than 2.8 $\mu$m so as to evaluate a real OH-band absorption.

\begin{figure}[ht]
\centering
\includegraphics[width=0.9\hsize]{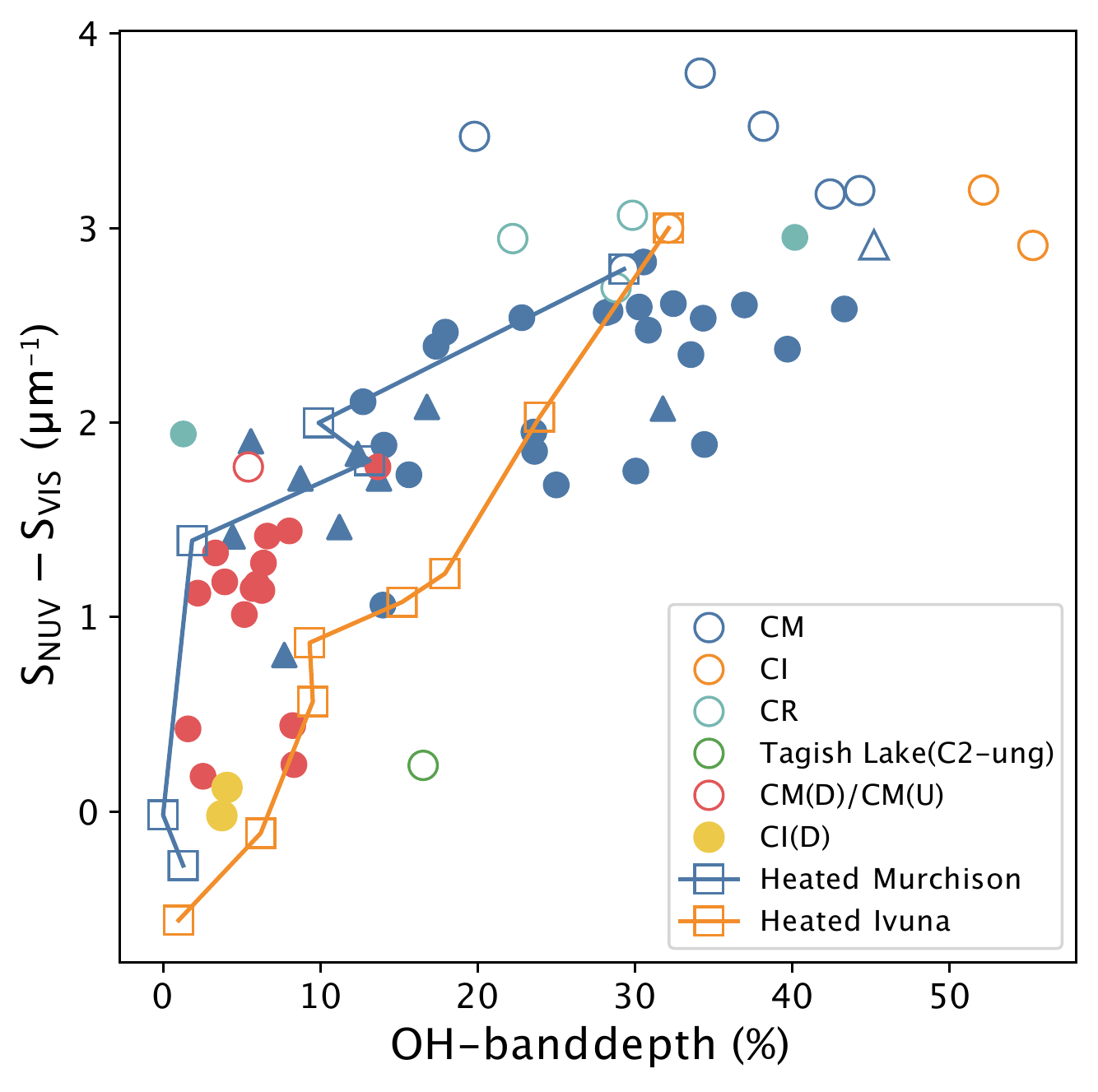}
\caption{OH-band depth and NUV absorption of hydrated carbonaceous chondritic meteorites. The meteorite sample spectra are from \citet{Hiroi2021} (available at the RELAB database). Open symbols indicate the powder samples and filled symbols indicate chip samples. CMs are classified into with (circle) and without (triangle) the 0.7-$\mu$m band absorption. Squares with lines show the heating experiments of Ivuna (CI1) and Murchison (CM2) conducted in \citet{Hiroi1996a,Hiroi1996b}. Unheated Ivuna and Murchison are shown by squares with circles inside. CM(D), CM(U), and CI(D) are ATCCs.}
\label{fig:met}
\end{figure}

\citet{Hiroi1993} has shown that naturally-heated CI/CM meteorites (ATCCs) and experimentally-heated Murchison samples have similar UV to NIR spectral shape. Their experimental and natural CI/CM spectra show the strong correlation between NUV and OH-band absorption strengths. However, at that time, the absorbed water was not taken into account for measurement of the OH-band and actually, a broad absorption around 3-$\mu$m can be observed in their spectra. Thus, using the given 6th-order Gaussian fitting by \citet{Hiroi2021}, we remove the contamination of terrestrial absorbed water by having into account the Gaussian components which have center wavelength shorter than 2.8 $\mu$m. Figure \ref{fig:met} shows the NUV and OH-band absorption strengths of hydrous meteorites. We find a similar correlation as in \citet{Hiroi1993}: the more hydrated meteorites show deeper NUV absorption. ATCC samples (CM(D)/CM(U)/CI(D)) follow well the heating experiment trends. This is because the phyllosilicates are decomposed by heating. Another important point here is that powder samples (open symbols) tend to show deeper absorptions than chip samples. When compared with other hydrated carbonaceous chondrites, Tagish Lake exhibits lower NUV absorption, possibly because absorption features are masked by high carbon contents, $\sim5.4$ wt\% \citep{Brown2000}. Tagish Lake has extremely low reflectance and does not have any feature in the wavelength range $<2.5$ $\mu$m. Tagish Lake is also known to have a very red visible reflectance spectrum \citep{Hiroi2001}. Tagish Lake’s dark-red spectrum without the strong NUV absorption suggest similarity with P and D classes. Some P-class asteroids exhibit OH-band absorption, suggesting that Tagish Lake could be the meteorite analog. Although a D-class asteroid with a clear OH-band absorption has not been found, the sample number is quite small and the analog meteorites of P or D classes need further investigation in OH-band region. Our result shows strong correlation between the NUV and the OH-band absorption strength in hydrous meteorite spectra. Especially, CMs and CIs exhibit up to 50\% of OH-band and 3 -- 4 $\mu$m$^{-1}$ NUV absorption strengths. CMs without 0.7-$\mu$m band absorption (blue triangles) exhibit relatively shallow absorptions in both OH-band and NUV.

\subsection{Asteroids}
\begin{figure}[ht]
\centering
\includegraphics[width=\hsize]{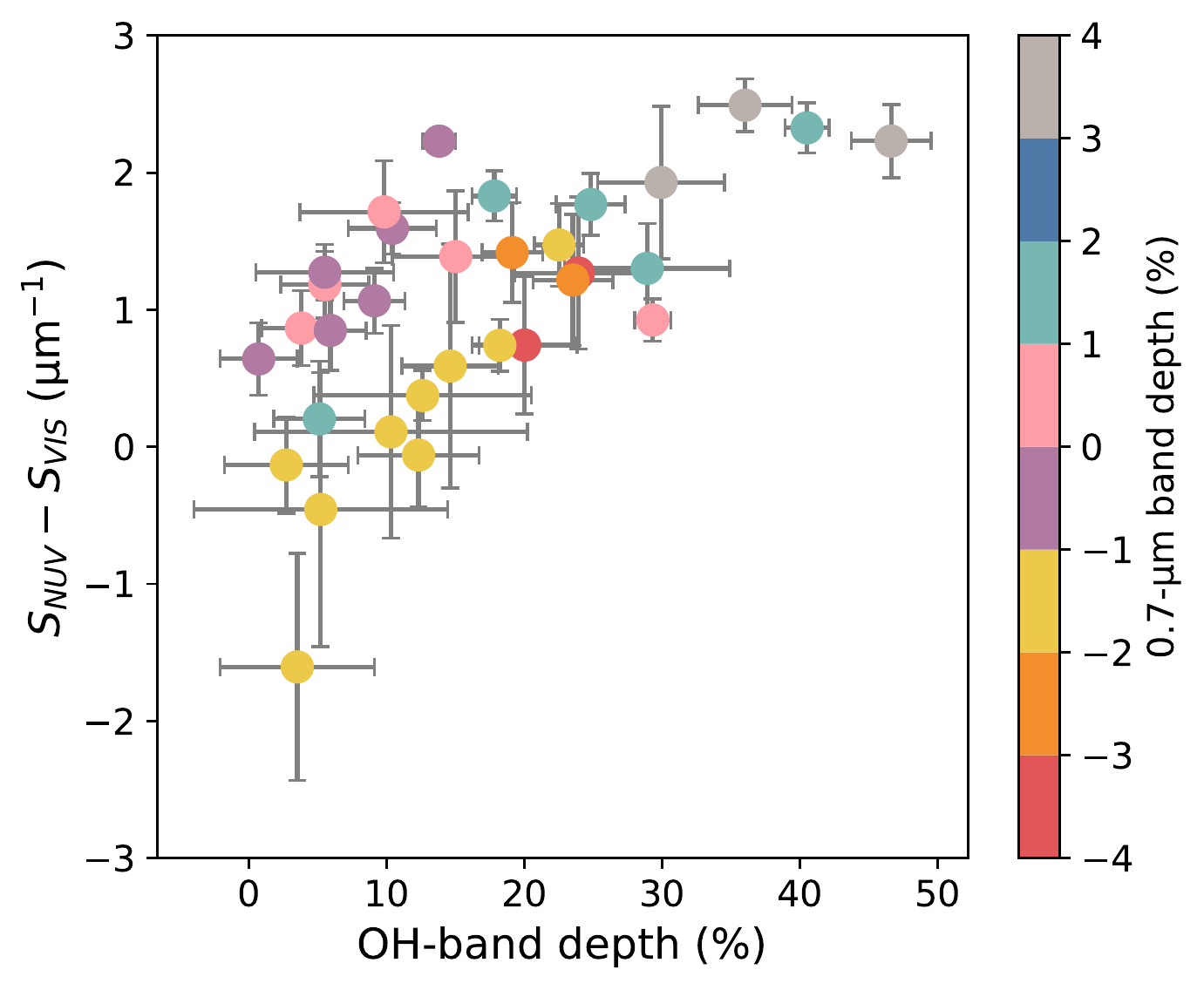}
\caption{OH-band depth and NUV absorption of primitive asteroids. Symbol color indicates the 0.7-$\mu$m band depth (HYD$_{\rm ECAS}$). Asteroids with deep 0.7-$\mu$m band absorptions show relatively deep OH-band and NUV absorptions. Even asteroids with the 0.7-$\mu$m band depth of $-1$\% to $1$\% exhibit slightly higher NUV absorptions compared with the ones with less 0.7-$\mu$m band depth.}
\label{fig:asteroid}
\end{figure}

AKARI is the infrared space telescope developed by JAXA \citep{Murakami2007}. Using the Infrared Camera onboard AKARI, infrared reflectance spectra from 2.5 to 5 $\mu$m of 66 asteroids were observed, including 34 
successfuly observed primitive asteroids (C-complex, P , and D classes) \citep{Usui2019}, making this the first large survey directly observing the 3-$\mu$m region. All the primitive asteroids observed by AKARI are larger than 80 km in diameter. 
Even though the correlation between OH-band and NUV absorption was suggested by \citet{Feierberg1985} and \citet{Hiroi1996a, Hiroi1996b}, at that time the OH-band (2.7-$\mu$m band) of asteroids was not directly observed and 2.9 -- 3.0 $\mu$m depth was measured instead. Thus, here we compare the 2.7-$\mu$m band depth from the AKARI data and corresponding NUV absorption from the ECAS data. 32 primitive asteroids were commonly observed by AKARI and ECAS. The 0.7-$\mu$m absorption using three bands of ECAS data, $v$, $w$, and $x$. Thus, the positive value of the 0.7-$\mu$m absorption means that the spectral shape is concave and shows an absorption, while the negative value of the 0.7-$\mu$m absorption means that the spectral shape is convex and has no clear absorption. Figure \ref{fig:asteroid} shows the positive correlation between OH-band depth and NUV absorption. Asteroids with NUV absorption > 2 $\mu$m$^{-1}$ show >30\% absorption in OH-band and >1\% absorption in the 0.7-$\mu$m band. Moreover, most asteroids with the 0.7-$\mu$m band > 1\% depth have NUV absorption > 1 $\mu$m$^{-1}$. It should be noted that even asteroids without the 0.7-$\mu$m band but the depth $>-1$\% (pink and purple circles) tend to show an offset towards deeper NUV absorption than those with depth $<-1$\% (yellow circles). This might be because their higher contents of Fe-rich phyllosilicates make the 0.7-$\mu$m region from convex to flatter. This is quite reasonable because both the 0.7-$\mu$m and NUV absorptions are possibly caused by iron in the phyllosilicates. Pink and purple circle ones may contain more Fe-rich phyllosilicates than yellow circle ones. Based on the correlation of OH-band and NUV absorptions observed both in asteroids and meteorites, we can estimate the degree of hydration based on the NUV absorption. 

\section{Distribution of NUV absorption in the asteroid main belt}\label{sec:newtax}
\begin{figure*}[ht]
\centering
\includegraphics[width=0.8\hsize]{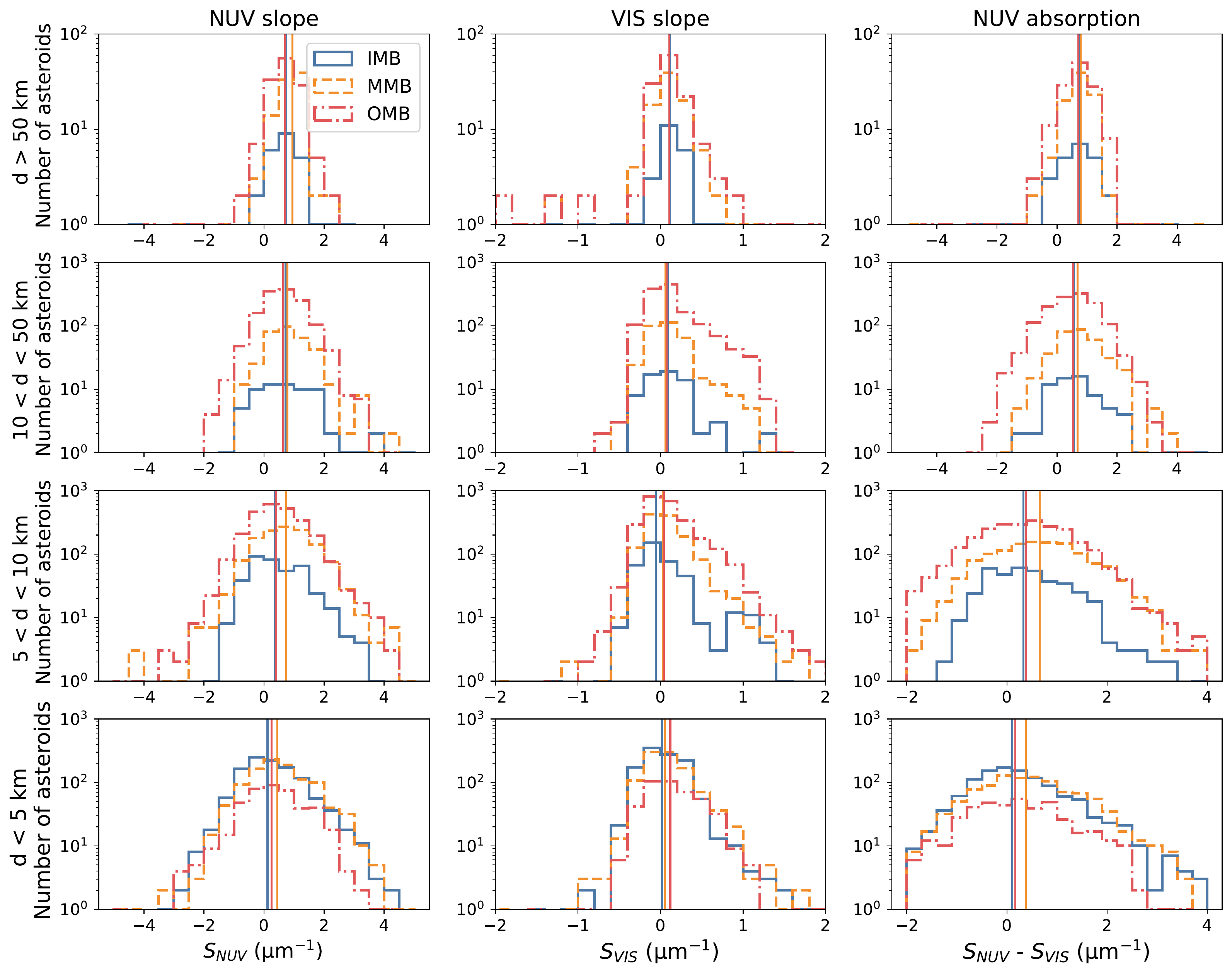}
\caption{Distribution of spectral slopes among the main asteroid belt. Left column: NUV slope, middle column: VIS slope, right column: NUV absorption strength ($S_{\rm NUV}-S_{\rm VIS}$). The Y axis indicates the number of asteroids in a bin. The blue line is the inner main belt, the yellow line is the middle main belt, and the red line is the outer main belt. The asteroids were divided by the diameter; first row: $d > 100$ km, second row: $50 < d < 100$ km, third row: $10 < d < 50$ km, fourth row: $5 < d < 10$ km. The vertical lines are the median value for each group.}
\label{fig:uvdist}
\end{figure*}

The asteroid main belt has been divided into three zones: inner (IMB, 2.06 -- 2.50 au), middle (MMB, 2.50 -- 2.82 au), and outer (OMB, 2.82 -- 3.28 au). Cybele and Hilda zones are located beyond the asteroid main belt at 3.3 -- 3.5 au (between 2:1 and 5:3 mean-motion resonances or MMR with Jupiter) and 4 au (the 3:2 MMR with Jupiter), respectively. Jupiter trojans are trapped in Jupiter's L4 and L5 lagragian regions at 5 au, so that they are considered to be dynamically stable. We also divided the asteroids into different size ranges by diameter $d$: very large ($d>100$ km), large ($50<d<100$ km), medium ($10 < d < 50$ km), small ($5<d<10$ km), very small ($d < 5$ km). The NUV absorption strength distributions for each region with different size ranges are shown in Fig. \ref{fig:uvdist}. The median value and interquartile range for each group is shown in Table \ref{table:uvdist}. To evaluate the difference in the distributions among IMB, MMB, and OMB asteroids, we calculated the $p$-values for each combination. First, we applied the Shapiro-Wilk test to test the normality for each group. If the $p$-value for a group is less than 0.05, indicating a non-parametric distribution, we used the Mann-Whitney U test to calculate the probability that the two groups of populations are equal. If the $p$-value of the Shapiro-Wilk test for a group is larger than 0.05, indicating a normal distribution, we used the Student’s t-test. For the large diameter populations ($d > 50$ km), the NUV absorption strengths are not significantly different ($p>0.05$ for all combinations IMB-MMB, IMB-OMB, and MMB-OMB). This suggests that the NUV distributions of the large diameter populations are similar through the main asteroid belt. On the other hand, the distributions of the small population ($d<10$ km) are different for IMB-MMB and MMB-OMB with $p \ll 0.01$, while the difference between IMB and OMB is not significant, $p>0.05$.  
Based on the statistical test, the IMB and OMB populations with $d<10$ km show less NUV absorption strengths (0.1 --0.4 $\mu$m$^{-1}$) compared with MMB (0.4--0.7 $\mu$m$^{-1}$). The distributions of medium and small populations in IMB and OMB are slightly different, $p<0.05$. The MMB distribution is similar between medium and small populations. Over all, MMB shows the deepest NUV absorption for all size groups except for the very large $d>100$ km group, suggesting that MMB are the most hydrated zone among the main asteroid belt. And IMB is least hydrated for $d < 10$ km.
%
%
%
\begin{table*}[ht]
\caption{NUV absorption strength for each size range and region in the main asteroid belt. IQR is the interquartile range of samples. We did not calculate IQR for $d>100$ km in IMB due to few number (5) of samples.}
\label{table:uvdist}
\centering
\begin{tabular}{c|cc|cc|cc|cc|cc}
\hline\hline
 & \multicolumn{10}{c}{$S_{\rm NUV}-S_{\rm VIS}$ ($\mu$m$^{-1}$)}\\
Region & \multicolumn{2}{c|}{$d>100$ km} & \multicolumn{2}{c|}{$50<d<100$ km} & \multicolumn{2}{c|}{$10<d<50$ km} & \multicolumn{2}{c}{$5<d<10$ km} & \multicolumn{2}{c}{$d<5$ km}\\
 & Med. & IQR & Med. & IQR & Med. & IQR & Med. & IQR & Med. & IQR\\ 
 \hline
 IMB & 1.35 & -- & 0.57 & 1.10 & 0.56 & 1.18 & 0.33 & 1.08 & 0.11 & 1.17\\
 MMB & 0.80 & 0.45 & 0.72 & 0.80 & 0.68 & 1.07 & 0.66 & 1.34 & 0.38 & 1.34\\
 OMB & 0.80 & 0.57 & 0.61 & 0.81 & 0.54 & 1.12 & 0.38 & 1.31 & 0.17 & 1.39\\
 All in MB & 0.80 & 0.54 & 0.63 & 0.93 & 0.57 & 1.12 & 0.44 & 1.32 & 0.21 & 1.28\\
 \hline
\end{tabular}
\end{table*}

\begin{figure}[ht]
\centering
\includegraphics[width=\hsize]{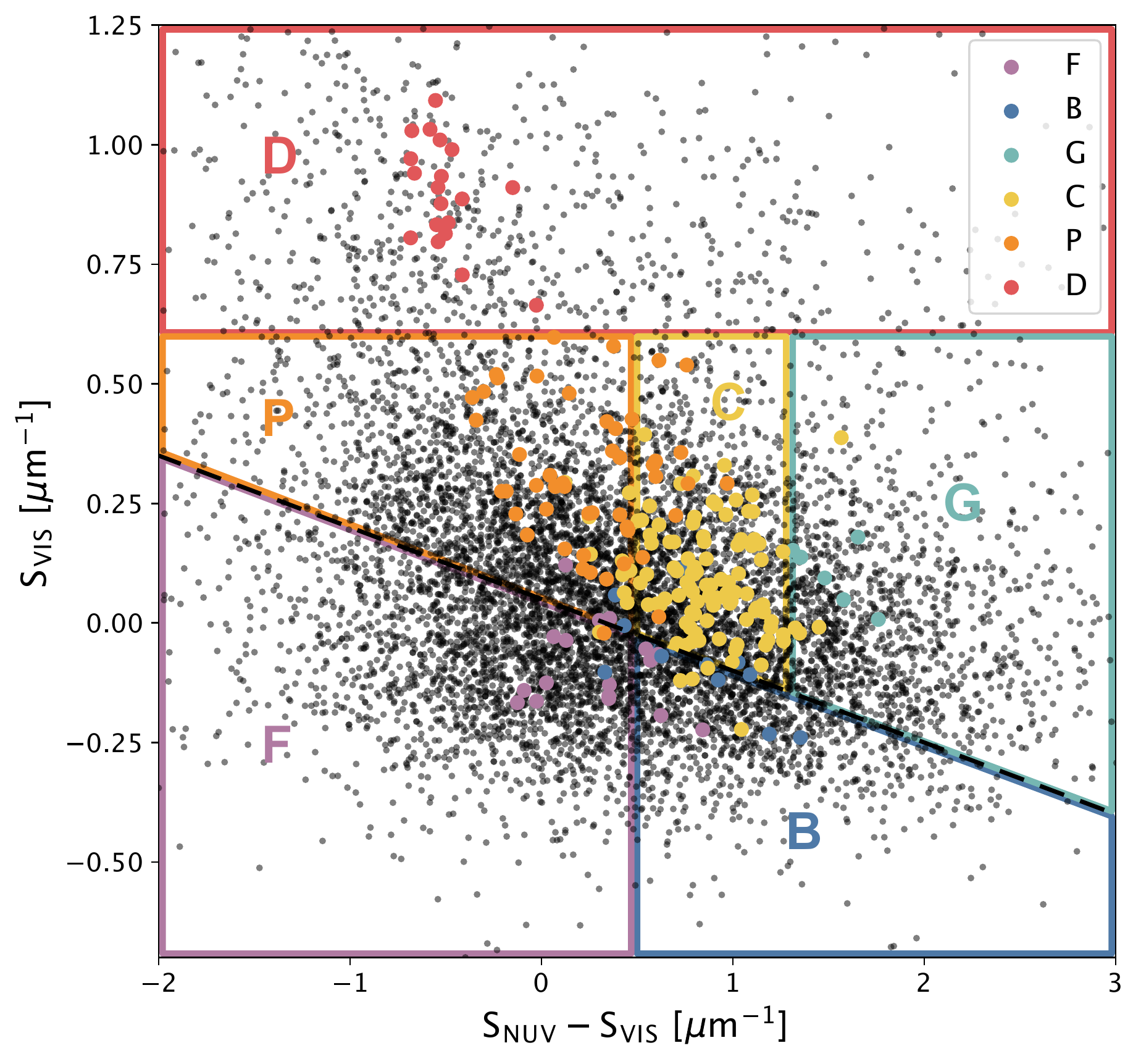}
\caption{Spectral characteristics of all primitive asteroids using the ECAS and SDSS datasets (black circles). The colored circles indicate the ECAS dataset with the taxonomy defined by \citet{Tholen1984}. The dashed line shows the direction of the largest dispersion of the black circles ($y=-0.15x+0.05$). The taxonomic classifications (Table \ref{table:threshold}) are also visualized.}
\label{fig:nuv-vis_tax}
\end{figure}

\begin{figure}[ht]
\centering
\includegraphics[width=\hsize]{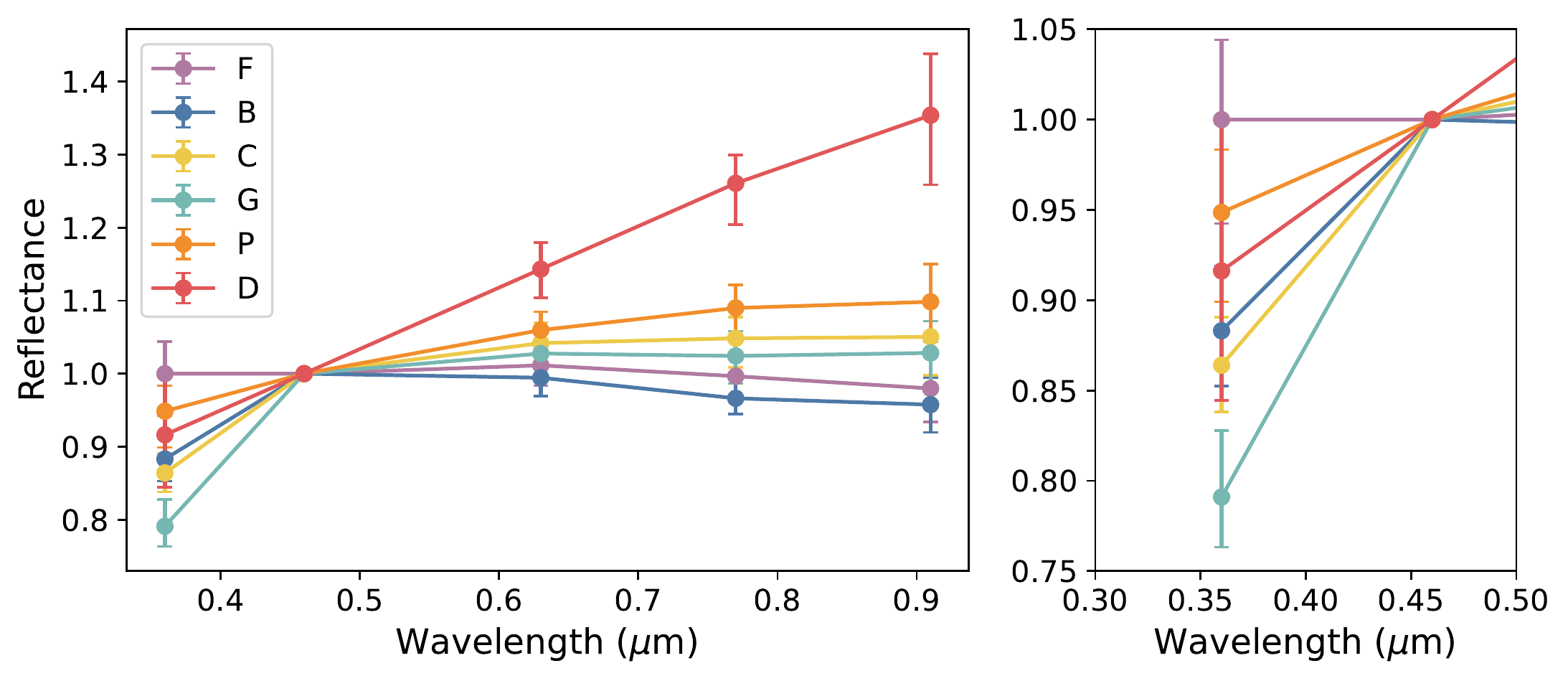}
\caption{Average spectrophotometry of taxonomy classes through the SDSS filter set classified by Table \ref{table:threshold}. The error bars show the interquartile range in the taxonomic classes. (left) The spectra for whole wavelength. (right) The same spectra closed up in the NUV.}
\label{fig:spectra_tax}
\end{figure}

\begin{figure*}[ht]
\centering
\includegraphics[width=\hsize]{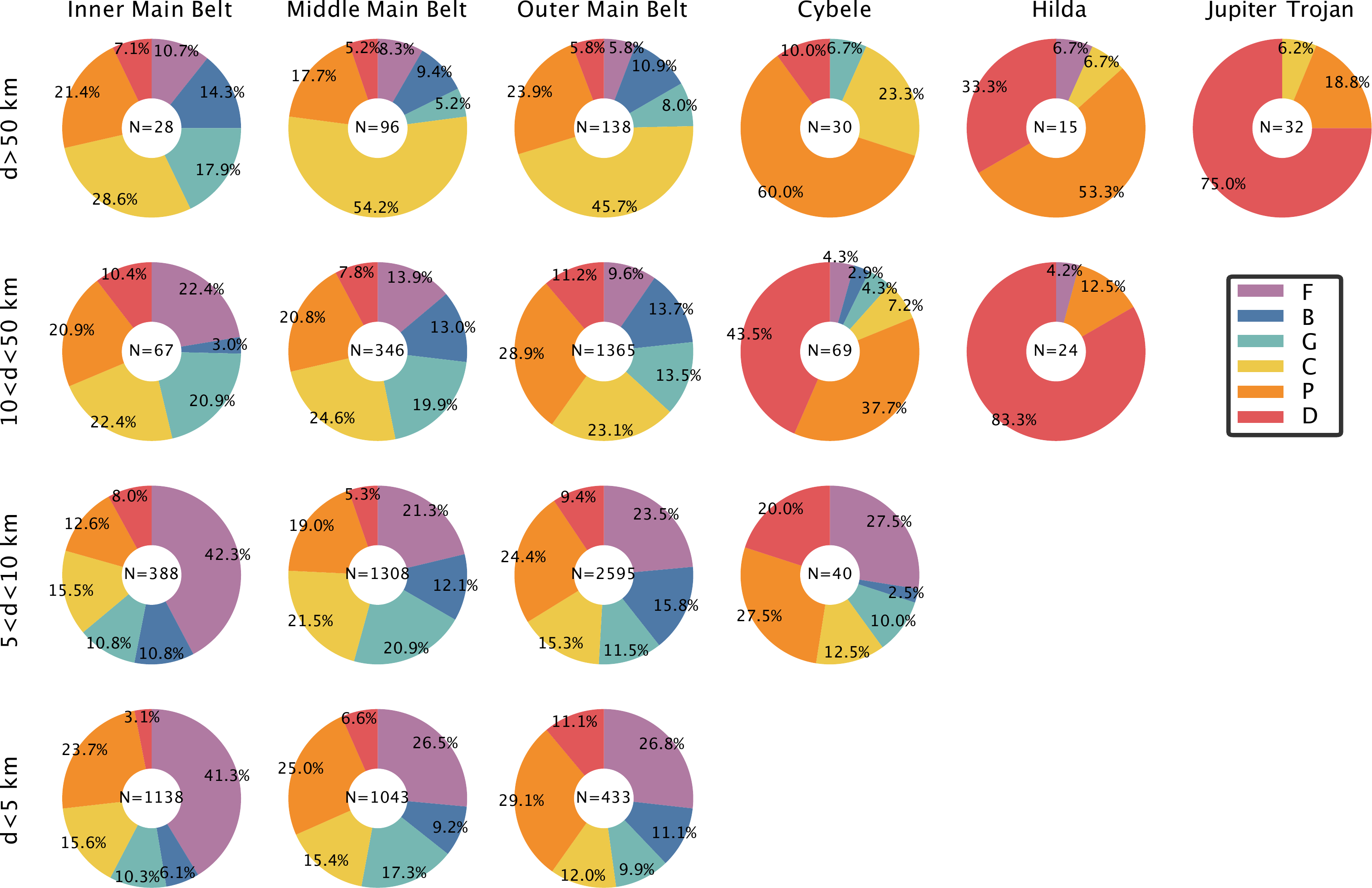}
\caption{Taxonomic distribution of primitive asteroids as a function of size and dynamical populations. The total number in each bin is shown in the center of pie chart.}
\label{fig:pi}
\end{figure*}

To depict the taxonomic distribution, we classified the primitive asteroids into F, C, B, G, P, and D using the values in Table \ref{table:threshold}. The thresholds for the classifications were defined based on the ECAS taxonomy. Figure \ref{fig:nuv-vis_tax} shows the distribution of our samples in the NUV absorption vs. VIS slope space. We depict the ECAS samples by colored circles in order to demonstrate the correspondence of our classification and the ECAS taxonomy. The line dividing blue, F and B classes, and red, C, G and P classes (L1) is defined by the direction of largest dispersion of dataset. The average spectrophotometry in the SDSS filters of taxonomy classified by Table \ref{table:threshold} are shown in Fig. \ref{fig:spectra_tax}. Using this classification, we could measure the number fraction of each taxonomy as a function of size and regions (Fig. \ref{fig:pi}). \citet{DeMeo&Carry2014} investigated similar distribution using only the VIS wavelength. Thus, they could not distinguish between C, F, and G, while our dataset including NUV region can distinguish them and exploit more compositional information (see Sect. \ref{sec:discussion}).

The C-type asteroids are dominant for asteroids $d> 50$ km diameter, and quite minor for smaller sizes. The F types are common throughout the main asteroid belt, even out to the small size of Cybele region. Components of the main belt are completely different between $d>50$ km and $d<10$ km and F type asteroids are the major component of $d<10$ km in IMB. All the classes are almost evenly distributed for MMB and OMB for small asteroids $d<10$ km, although the enhancement of F class is found for the asteroids $d<10$ km. We also note that P-class asteroids are > 20\% through most of the main asteroid belt regions and size ranges, which is comparable to other types, suggesting that P types are not a minor component of the main asteroid belt. Although P types are dominant for the Cybele and Hilda regions, if P types originated from the outer Solar System, they have significantly contaminated the main asteroid belt. Nevertheless the abundance of P types through the main asteroid belt suggests that P types were formed in the same reservoir as the C complex asteroids. In contrast, in most main asteroid belt zones and sizes, the D-type asteroids are minor $\leq$ 10\%, while they are a main component for the large size Jupiter Trojans, and the middle size Hilda and Cybele zones. The Cybele and Hilda zones show completely different taxonomic population from size to size, which agree with the result form \citet{DeMeo&Carry2014}. In the Cybele zone P, and D types are dominating for population $d > 10$ km, whereas F types are the largest population in $d < 10$ km.

Our distribution can be compared with \citet{Tholen1984}. As most asteroids in \citet{Tholen1984} is larger than 50 km, we can compare simply the distribution of $d>50$ km. \citet{Tholen1984} found that the C types are dominant among primitive asteroids in the main belt: $\sim 70$\% for IMB, $\sim 80$\% for MMB, and $\sim 70$\% for OMB. Our result also suggests the dominant distribution of C types for large asteroids, but not as much as \citet{Tholen1984}. Instead, we found more P types distributing through the main belt, $\sim 20$\%, which is consistent with \citep{DeMeo&Carry2014}. This is may be because P types have relatively low albedo and could be missed by the discovery bias. Similarly, we also found several D types in the main belt. The distributions in IMB show a quite difference between our result and \citet{Tholen1984} may be because of the sample number of \citet{Tholen1984} is less than 10.

Beyond the main belt, the rapid decrease of C type as suggested by \citet{Tholen1984}. In the Cybele zone, \citet{Tholen1984} found $\sim 35$\% for both C and P types and $\sim 20$\% of D type, while our result suggest more abundance of P type in large asteroids and D type is only 10\%. Our result is more consistent with \citep{DeMeo&Carry2014}, and again this could be the sample bias. In the Hilda and Jupiter Trojan zones, our result is quite consistent with \citet{Tholen1984} because these populations are dominated by P and D types, which both have comparable low albedos.
\begin{table}
\caption{Thresholds for taxonomic classification used in this study to classify asteroids similar to Tholen's taxonomy. L1 is the line dividing bluer (F and B) and redder objects in VIS. L1 is defined as the line along the largest dispersion in the dataset.}
\label{table:threshold}
\centering
\begin{tabular}{c c c cc}
\hline\hline
Class & \multicolumn{2}{c}{$S_{\rm VIS}$ ($\mu$m$^{-1}$)} & \multicolumn{2}{c}{$S_{\rm NUV}-S_{\rm VIS}$ ($\mu$m$^{-1}$)}\\
 & min & max & min & max \\ \hline
 C & L1 & 0.25 & 0.5 & 1.1 \\
 F & -- & L1 & -- & 0.5 \\
 G & L1 & 0.25 & 1.1 & -- \\
 B & -- & L1 & 0.5 & 1.1 \\
 P & L1 & 0.6 & -- & 0.5 \\
 D & 0.6 & -- & -- & --\\
 \hline
\end{tabular}
\end{table}
\begin{table*}[ht]
\caption{Primitive asteroid distribution over taxonomic classes and dynamical populations for different asteroid size bins.}
\label{table:matome}
\centering
\begin{tabular}{crrrrrrrr}
\multicolumn{2}{l}{All} \\
\hline\hline
Class & Samples & IMB & MMB & OMB &  Cybele &  Hilda & Trojan & Others\\ \hline
 F & 2,154 & 652 & 610 & 866  & 15 & 2 & 0 & 0\\
 B & 1,089 & 117 & 308 & 661  & 3  & 0 & 0 & 0\\
 G & 1,252 & 178 & 528 & 537  & 9  & 0 & 0 & 0\\
 C & 1,687 & 260 & 579 & 827  & 18 & 1 & 2 & 0\\
 P & 2,197 & 339 & 598 & 1,187 & 55 & 11 & 7 & 0\\
 D & 798   & 75  & 170 & 453   & 11 & 25  & 29 & 4\\ \hline
 Total & 9,168 & 1,621 & 2,793 & 4,531 & 142 & 39 & 38 & 4\\
 \hline
\end{tabular}

\vspace{3mm}
\begin{tabular}{crrrrrrrr}
\multicolumn{2}{l}{$d> 50$ km} \\
\hline\hline
Class & Samples & IMB & MMB & OMB &  Cybele &  Hilda & Trojan & Others\\ \hline
 F & 20 & 3 & 8 & 8 & 0 & 1 & 0 & 0\\
 B & 28 & 4 & 9 & 15 & 0 & 0 & 0 & 0\\
 G & 23 & 5 & 5 & 11 & 2 & 0 & 0 & 0\\
 C & 133 & 8 & 52 & 63 & 7 & 1 & 2 & 0\\
 P & 88 & 6 & 17 & 33 & 18 & 8 & 6  & 0\\
 D & 50 & 2  & 5 & 8 & 3 & 5 & 24  & 3\\ \hline
 Total & 342 & 28 & 96 & 138 & 30 & 15 & 32 & 3\\
 \hline
\end{tabular}

\vspace{3mm}
\begin{tabular}{crrrrrrrr}
\multicolumn{2}{l}{$10 <d< 50$ km} \\
\hline \hline
Class & Samples & IMB & MMB & OMB &  Cybele &  Hilda & Trojan & Others\\ \hline
 F & 198 & 15 & 48 & 131 & 3 & 1 & 0 & 0\\
 B & 236 & 2 & 45 & 187 & 2 & 0 & 0 & 0\\
 G & 270 & 14 & 69 & 184 & 3 & 0 & 0 & 0\\
 C & 420 & 15 & 85 & 315 & 5 & 0 & 0 & 0\\
 P & 511 & 14 & 72 & 395 & 26 & 3 & 1  & 0\\
 D & 243 & 7  & 27 & 153 & 30 & 20 & 5  & 1\\ \hline
 Total & 1,878 & 67 & 346 & 1,365 & 69 & 24 & 6 & 1\\
 \hline
\end{tabular}

\vspace{3mm}
\begin{tabular}{crrrrrr}
\multicolumn{2}{l}{$5 <d< 10$ km} \\
\hline\hline
Class & Samples & IMB & MMB & OMB &  Cybele\\ \hline
 F & 1,064 & 164 & 278 & 611 & 11 \\
 B & 612  & 42  & 158  & 411 & 1  \\
 G & 619  & 42  & 274  & 299 & 4  \\
 C & 743  & 60  & 281  & 397 & 5  \\
 P & 941  & 49  & 248  & 633 & 11 \\
 D & 352 & 31   & 69   & 244 & 8  \\ \hline
 Total & 4,331 & 388 & 1,308 & 2,595 & 40 \\
 \hline
\end{tabular}
\hspace{5mm}
\begin{tabular}{crrrrrr}
\multicolumn{2}{l}{$d< 5$ km} \\
\hline\hline
Class & Samples & IMB & MMB & OMB &  Cybele\\ \hline
 F & 863 & 470 & 276  & 116 & 1 \\
 B & 213  & 69  & 96  & 48 & 0  \\
 G & 340  & 117  & 180 & 43 & 0  \\
 C & 391  & 117  & 161  & 52 & 0  \\
 P & 657  & 270  & 261  & 126 & 0 \\
 D & 153 & 35   & 69   & 48 & 1  \\ \hline
 Total & 2,617 & 1,138 & 1,043 & 433 & 3 \\
 \hline
\end{tabular}
\end{table*}

\section{Near-infrared characteristics of the primitive asteroids}\label{sec:movis}
The MOVIS catalogs \citep{Popescu2016, Popescu2018} include the surveys of Solar System objects observed  in a serendipitous manner by VISTA-VHS (Visible and Infrared Survey Telescope for Astronomy - VISTA Hemisphere Survey) program \citep{McMahon2013, Sutherland2015}. This survey covered the entire southern sky hemisphere by using the near-infrared broad-band filters $Y$ (centered at 1.020 $\mu$m), $J$ (1.252 $\mu$m), $H$ (1.645 $\mu$m), and $Ks$ (2.147 $\mu$m). The data retrieved for Solar System objects include: the detections catalog (MOVIS-D), the magnitudes catalog (MOVIS-M), and the colors catalog (MOVIS-C). The number of measured colors for each Solar System object varies according to the observing strategy and to the limiting magnitude of each filter. The average time interval between the measurements (which affect the color uncertainty) done with $Y$ and $J$ filters is $8.47~\pm~5.99$ min, while  for $J$ and $Ks$ filters is $7.40~\pm~6.56$ min. This time interval is negligible compared with the typical rotation period of main belt asteroids, which is in the order of several hours \citep{Warner2009}, thus the errors introduced by the lightcurve variations can be ignored.
 
\begin{table}[ht]
\caption{The $(Y-J)$ and $(J-Ks)$ of each taxonomic class discussed in this work. The average values, the standard deviation ($\sigma$), and the median values are shown.}
\label{table:nircolors}
\vspace{3mm}
\begin{tabular}{c| c c c| c c c}
\hline \hline
 Class &       & $(Y-J)$  &       &  &  $(J-Ks)$ & \\ \hline
      & avg.  & $\sigma$ & med.  &  avg.  &  $\sigma$ & med.\\ \hline
 D    & 0.360 & 0.135 & 0.346 & 0.538 & 0.102 & 0.563\\
 P    & 0.278 & 0.061 & 0.279 & 0.485 & 0.120 & 0.490\\
 C    & 0.258 & 0.040 & 0.252 & 0.421 & 0.076 & 0.420\\
 B    & 0.238 & 0.025 & 0.242 & 0.467 & 0.075 & 0.478\\
 F    & 0.244 & 0.025 & 0.241 & 0.420 & 0.070 & 0.427\\
 G    & 0.261 & 0.064 & 0.258 & 0.410 & 0.070 & 0.427\\
 \hline
\end{tabular}
\end{table}

We used the latest version of the MOVIS-C catalog \citep{Popescu2018} to compare with the results we obtained based on the SDSS and ECAS dataset. We found 242 objects in common with accurate $(Y-J)$ and $(J-Ks)$ colors, where respectively the errors follow the conditions $(Y-J)_{\rm err} \leq 0.05$ and $(J-Ks)_{\rm err} \leq 0.05$. This sample includes 39 samples classified as D type, 79 P types, 40 C types, 21 B types, 31 F types, and 32 G types based on our classification in Sect. \ref{sec:newtax}. Table \ref{table:nircolors} shows some statistical parameters of $(Y-J)$ and $(J-Ks)$ colors for each of these classes.

The differences between C, P and D-types are outlined in both $(Y-J)$ and $(J-Ks)$ colors and follows the reddening trend from C $\rightarrow$ P $\rightarrow$ D types.  The average $(J-Ks)$ color for P is $(J-Ks)_{\rm P}=0.485\pm0.120$ mag which is almost 1$\sigma$  larger than the average of C types,  $(J-Ks)_{\rm C}=0.421\pm0.076$  mag. In this wavelength region (1.25 -- 2.2) $\mu$m,  P types have a broad spectral behavior (outlined by the value of dispersion of its colors). The $(Y-J)$ color distributions of C and P types are slightly different, although the average values are comparable, $(Y-J)_{\rm C}=0.258\pm0.040$ and
$(Y-J)_{\rm P}=0.278\pm0.060$. The marginally red spectrophotometry (over 1.020 -- 1.252 $\mu$m  of these two types can be inferred by comparing these average values with the median  for the solar analogs $(Y-J)_{\rm G2V}=0.219$ \citep{Popescu2018}.
 
\begin{figure*}[ht]
\centering
\includegraphics[width=6cm,height=4.5cm]{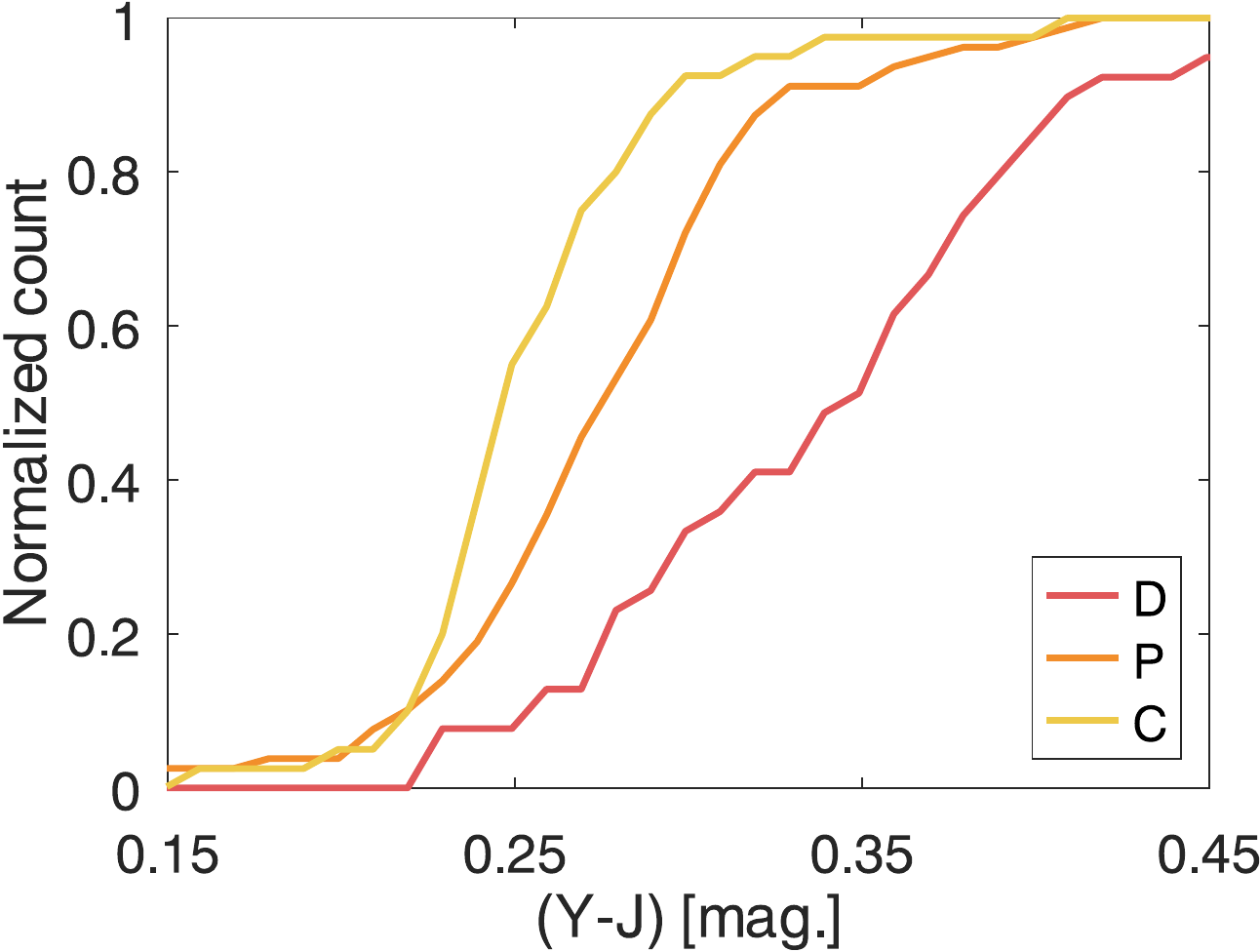}
\includegraphics[width=6cm,height=4.5cm]{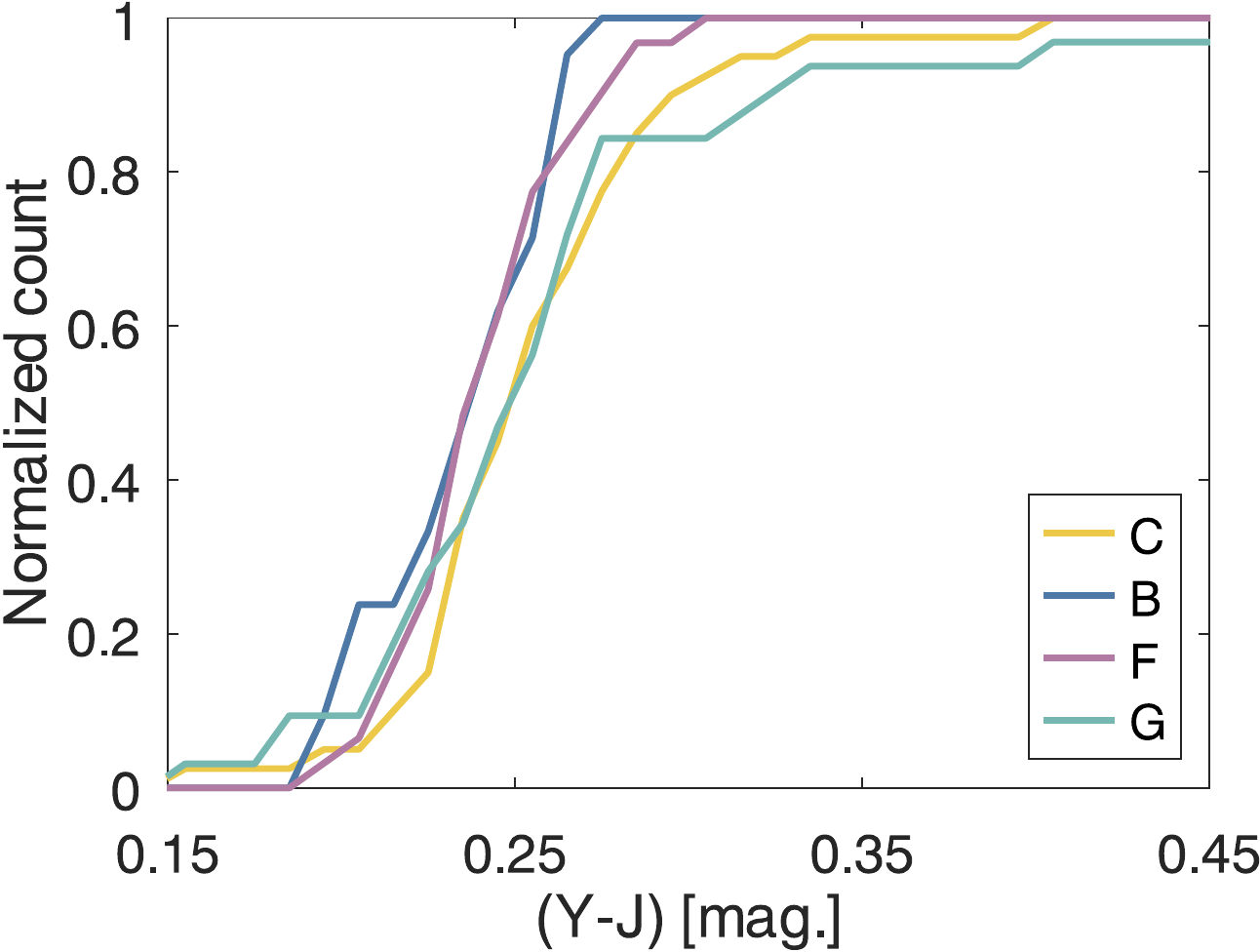}\\
\includegraphics[width=6cm,height=4.5cm]{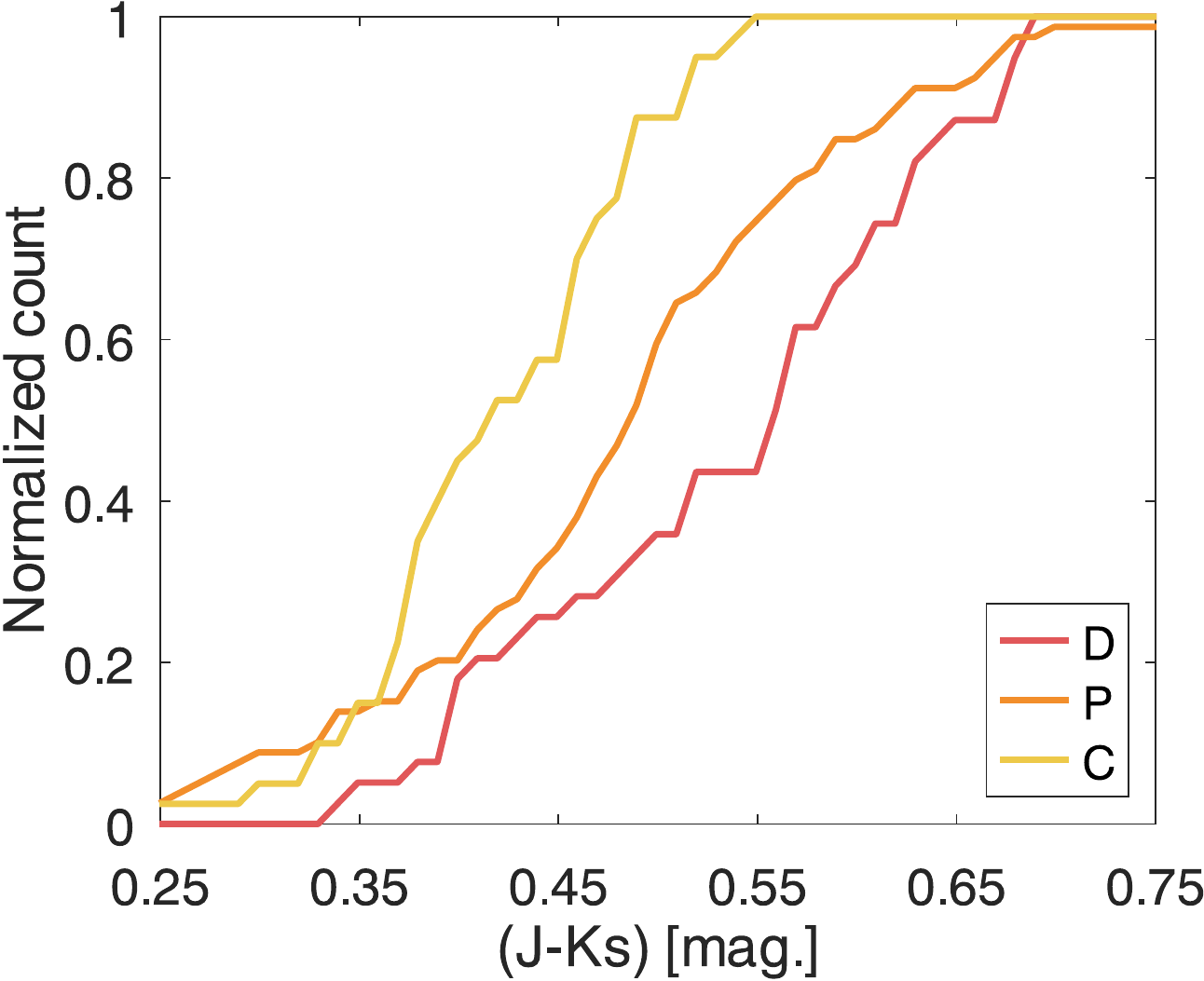}
\includegraphics[width=6cm,height=4.47cm]{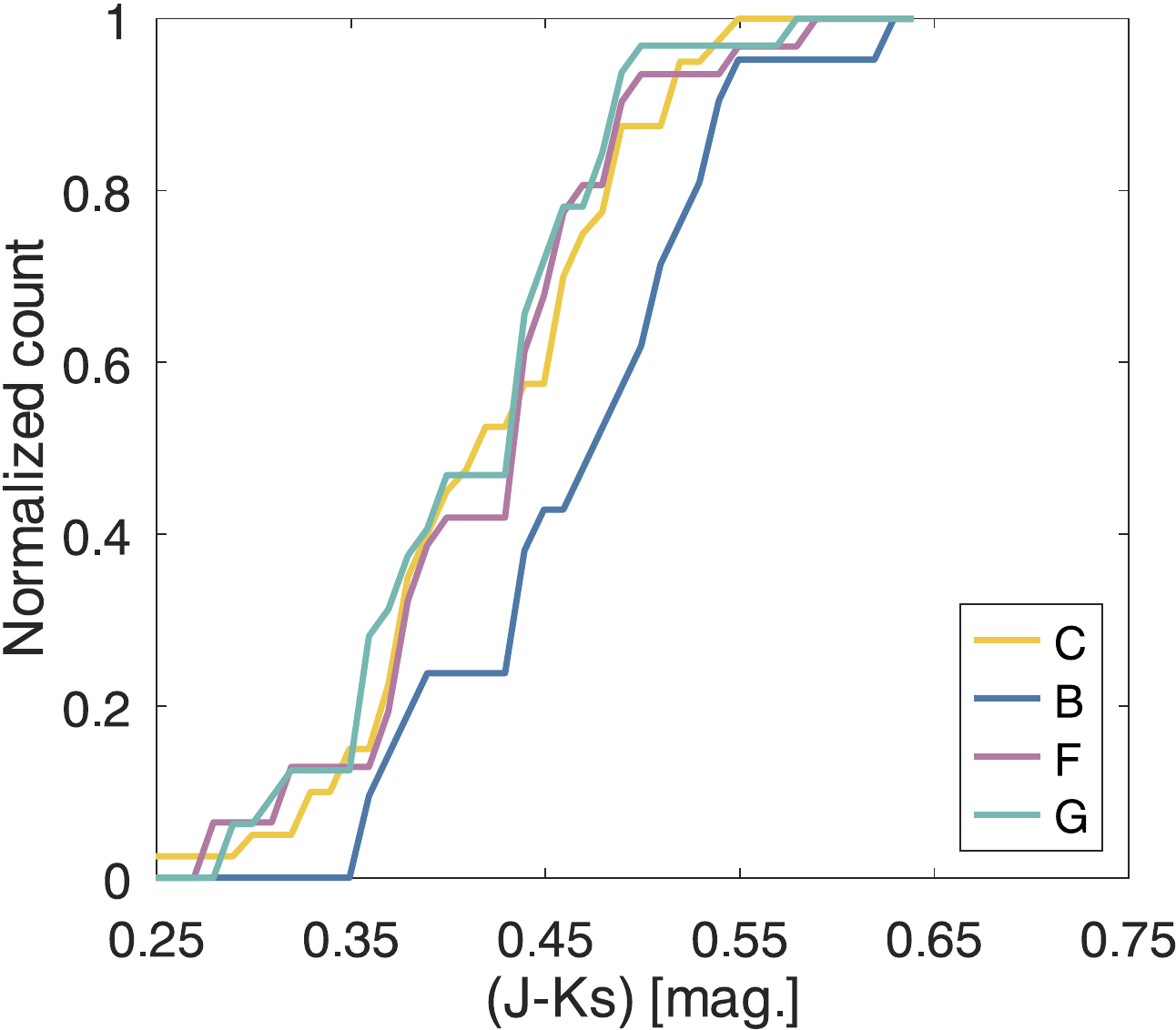}
\caption{Normalized cumulative distribution of $(Y-J)$ (top panels), and of $(J-Ks)$ (bottom panels) for the sample of 242 objects taxonomically classified by this work using the ECAS and SDSS dataset.}
\label{fig:MOVIS}
\end{figure*}

D types have a well separated distribution for both $(Y-J)$ and $(J-Ks)$ color (Fig. \ref{fig:MOVIS}). The average values are very red: $(Y-J)_{\rm D}=0.360\pm0.135$ mag (a value about 2$\sigma$ larger compared to the rest of primitive classes), and $(J-Ks)_{\rm D}=0.537\pm0.102$ mag. The wide spread of $(J-Ks)$  may indicate compositional variations inside this group.

The $(Y-J)$ color distribution of the B, C, F, and G classes are similar (Table \ref{table:nircolors}). The color values are smaller than those of P and D classes. 
We found that B types exhibit red spectral spectrophotometric slope, indicated by their $(J-Ks)$ color (Table \ref{table:nircolors}). This shows potential differences in NIR. This spectral turning up at longer wavelengths 1.2 -- 2.1 $\mu$m might be similar to the trend in the Themis group classified by \citet{Clark2010} and the G1 and G2 groups in \citet{deLeon2012}. They found a good match between the B-type asteroids with this NIR upturn and CM, CI, and thermally metamorphosed CI/CM. 
We note that the B types defined by \citet{DeMeo2009}  should have negative NIR slopes, which means that $(Y-J)$ and $(J-Ks)$ color should have a lower value compared with the colors of solar analogs which are $(Y-J)_\odot = 0.219$ mag, $(J-Ks)_\odot = 0.336$ mag \citep{Popescu2018}. However, our classification was made using only the NUV-VIS colors provided by the SDSS filters applying Tholen's taxonomy. 

Figure \ref{fig:MOVIS} outlines that less than $\approx10\%$ of our C, B, F, and G types are below this threshold, meaning DeMeo's B type. Thus, majority of the B, C, F, and G types found in this sample have positive spectral slopes in the 1.02 -- 2.2 $\mu$m spectral intervals. 
This can be explained by considering the spectral variety presented by \citet{Clark2010} and \citet{deLeon2012}. They found that asteroids classified as B types according to their visible spectra show a broad variation of NIR spectral slopes, ranging from negative, blue slopes, to positive, moderately red slopes. The strong correlation between the NUV-VIS and NIR wavelengths is not found in the C-complex classes, i.e., B, C, F, and G classes, while B types may exhibit a potential difference in NIR. 

\section{Discussion}\label{sec:discussion}
\subsection{Aqueous alteration through the asteroid belt}
It is important to reveal the distribution of the aqueous alteration state through the current asteroid belt to constrain the boundary conditions for the Solar System evolution calculation such as the Grand-Tack model \citep[e.g.,][]{Walsh2011} and the Nice model \citep[e.g.,][]{Morbidelli2005}. We here discuss the aqueous alteration state as a function of the semi-major axis based on NUV, 0.7-$\mu$m band, and 3-$\mu$m features.

Based on the OH-band depth by indirect measurements of ground-based observations of a reflectance drop from 2.5 to 2.9 $\mu$m, \citet{Jones1990} found that the hydrated silicates slightly decline in abundance with distance from the Sun from 2.5 to 3.5 au because of the P/D population in the outer main belt. However, more recent work by \citet{Rivkin2015} argue that when focusing on the Ch asteroids, OH-band depth is poorly correlated with the semi-major axis, suggesting the population is well mixed in the asteroid belt. Direct measurements of the OH-band depth by the AKARI space telescope did not find the correlation between OH-band depth and the current semi-major axis of asteroids, either \citep{Usui2019}. Note that the asteroids which were observed in this region are quite large in size, i.e., more than several tens of km. Thus, by looking at the large members the hydrated state does not have clear correlation with the semi-major axis through the main belt. This is consistent with the similar NUV absorption strength through the main asteroid belt (Fig. \ref{fig:uvdist}).

Nevertheless, based on the band shape, the broad and rounded 3.1-$\mu$m band, possibly ice frost, was found in semi-major axes beyond 3.1 au coupled with the P and D types, while the sharp 3.1-$\mu$m band was found in a very wide range of 2.5 -- 4 au \citep{Takir&Emery2012}. This may point that the outer main belt might preserve the composition of anhydrous silicates, water ice, and possibly complex organic materials originated from the outer Solar System.

The 0.7-$\mu$m band and the OH-band features are highly correlated, suggesting the presence of the 0.7-$\mu$m band indicates phyllosilicates resulting from the aqueous alteration process \citep{Vilas1994, Howell2011, Fornasier2014, Usui2019}. Using the ECAS dataset, the distribution of 0.7-$\mu$m absorption was investigated by \citet{Vilas1994}, finding that the percentage of objects showing 0.7-$\mu$m absorption in different spectral types decreases from G > C > B > F > P > D. \citet{Fornasier2014} came to the same conclusion using spectroscopic observations. Asteroids with 0.7-$\mu$m band were proposed to dominate a zone from 2.6 to 3.5 au by \citet{Vilas1994} and later on it was updated to a zone from 2.3 to 3.1 au by \citet{Fornasier2014}. Furthermore, \citet{Fornasier2014} found that more than half of the objects in the IMB show the 0.7-$\mu$m band although MMB is the main region of hydrated objects in terms of number among the main asteroid belt. \citet{Rivkin2012} tried to measure the percentage of objects with 0.7-$\mu$m band using the SDSS photometry data. They found a percentage of 30\% to 40\% through the main asteroid belt although there are slight differences: OMB is the lowest, MMB is the highest. To summarize the observation of the 0.7-$\mu$m band, the MMB has the highest percentage of objects with the 0.7-$\mu$m band absorption.

The NUV absorption strength decreases from G > C/B > F > P > D, which is in good agreement with the percentage of objects showing 0.7-$\mu$m band found by \citet{Vilas1994} and \citet{Fornasier2014}. The G-type asteroids, the best analogs of the CM chondrites, characterized by NUV and 0.7-$\mu$m band absorptions, are distributed in the MMB with highest ratio. The B types are possibly analogs of CI or CM considering the turning up in NIR. Most B types distribute in MMB and OMB. The C types show moderate NUV absorption and almost flat reflectance spectra at VIS-NIR. The C types are prominent in the large members of main belt, suggesting possible survived primordial bodies from beginning. Thus, the C type spectra might show the composition of the outer most surface, such as NH$_4$-bearing phyllosilicates observed on the surface of (1) Ceres \citep{Ammannito2016}. The P/D types are also abundant in MMB, OMB, and beyond, although the domination in the Cybele, Hilda, and Jupiter Trojan zones is remarkable. However, P types are more abundant in the main belt region compared with D types. P types might be an end-member of C complex asteroids, which agrees with the previous studies \citep{Vernazza2017, Vernazza2021, Mahlke2022}. D types are hypothesized to be implanted in the Cybele and Hilda regions later than the formation of the main belt \citep{Levison2009}. 

In contrast, the IMB, especially in small members, is dominated by F types, the least NUV absorption strength among C complexes, which could be composed of Fe-poor phyllosilicates like CI chondrites or thermally metamorphosed carbonaceous chondritic materials. 
Previously, the F types were hypothesized as thermally metamorphosed carbonaceous chondrites due to their little NUV absorptions \citep{Hiroi1993}. If this is the case, the asteroids in IMB might have experienced hight temperature when formed, e.g., rapid formation and/or large-size planetesimals \citep[e.g.,][]{Grimm&Mcsween1993, Neumann2020}. It also should be noted that small asteroids could be fragments of catastrophic disruptions and have rubble-pile structure, and thus they may exhibit the internal compositions. This proposes the possibility that asteroids currently in IMB had relatively less heated crust and highly heated internal composition. Alternatively, they could be CI chondritic material (see also Sect. \ref{sec:ryugu}). This CI chondritic material could originated from fragments of core of large $d>100$ km IDP-like bodies as hypothesized by \citet{Vernazza2017}. In that case, they experienced low heating and might be formed in outer Solar System, e.g., slow formation and/or high water-rock ratio \citep[e.g.,][]{Nakamura2022}. These two hypotheses for the F type composition will inspire totally different history of the Solar System formation. However, we do not have clear evidence to conclude. Thus, more sample return missions will have importance. If the second case is plausible, the unique distribution of F types suggests that they were implanted in their current positions by different mechanisms apart from other primitive asteroids, or they were formed in a distinctive place in the early Solar System.

Summarizing, the characteristics of aqueous alteration through the main belt to the Cybele and Hilda zones can be depicted as 1) Fe-poor phyllosilicates or thermally metamorphosed carbonaceous chondrites in asteroids with $d<10$ km of IMB, 2) Fe-rich phyllosilicates in asteroids with $d>10$ km of MMB and OMB, 3) none or little aqueously altered comet-like materials with water frost in the Cybele, Hilda, and Jupiter Trojan zones, and 4) partially aqueously altered materials for all regions.  
\begin{figure}[ht]
\centering
\includegraphics[width=\hsize]{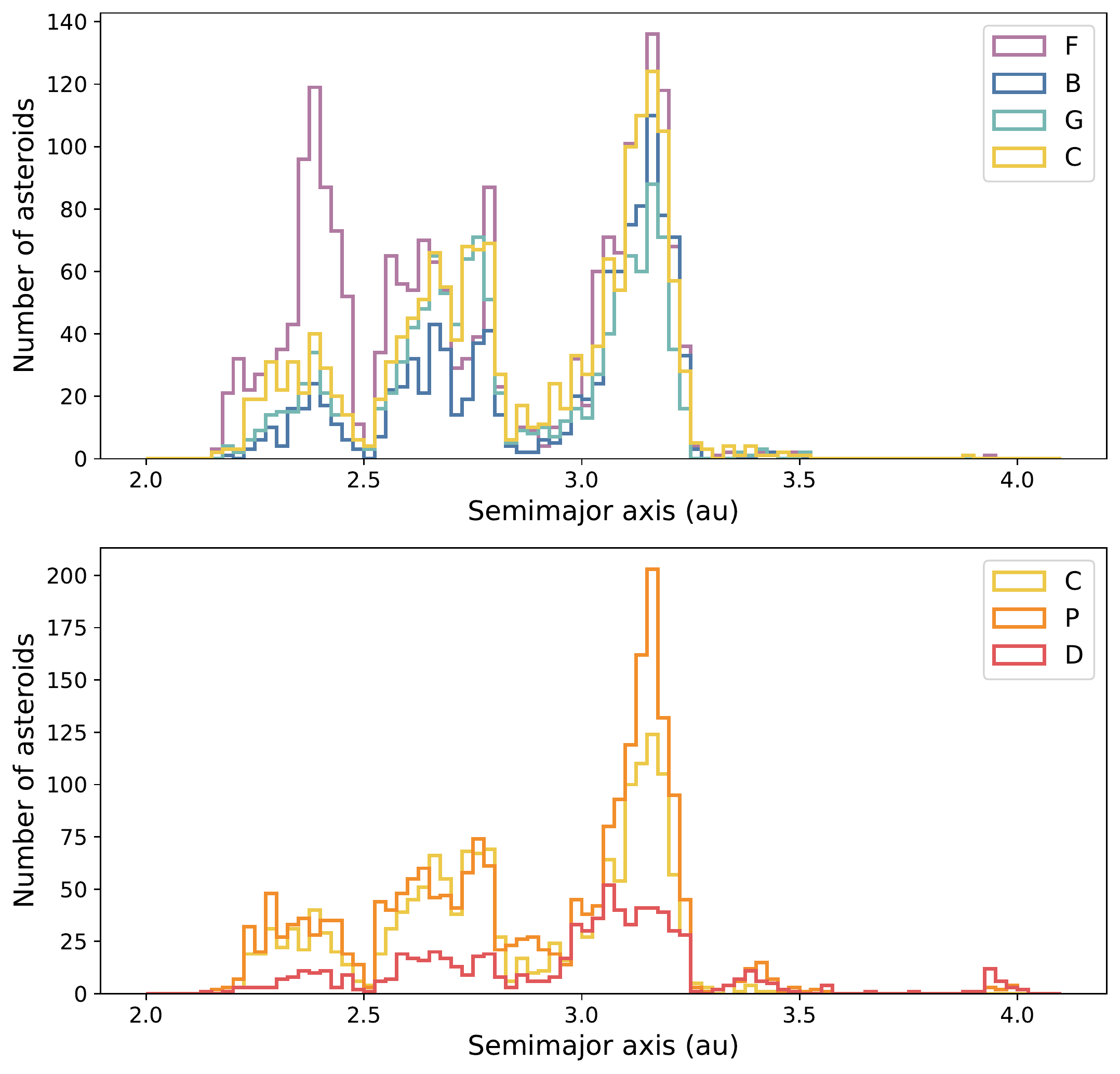}
\caption{Taxonomy distribution of primitive asteroids through the main belt to the Cybele and Hilda zone; the distributions of C, F, G, and B types (top) and those of C, P, and D types (bottom).}
\label{fig:semimajor_tax}
\end{figure}

\subsection{Grain size and surface physical condition}
It is known that the physical surface conditions affect the spectral slopes. Fine grain samples show redder spectral slope over VIS to NIR wavelengths \citep[e.g.,][]{Cloutis2018}. Grain size of asteroid surfaces can be estimated through thermal inertia derived by thermal infrared observations. Thermal inertia is indicative of surface roughness at a scale bigger than the typical diurnal heat propagation distance. Thermal inertia correlates with surface grain size: high thermal inertia can be interpreted as indicating coarse or rigid rock surface and low thermal inertia as indicating fine grain regolith \citep{Gundlach&Blum2013}. Based on the thermal infrared observations, increasing thermal inertia with decreasing diameter of the asteroid was found \citep{Delbo2007, Hanus2018}. Specifically, many of the asteroids with $d > 10$ km show low thermal inertia of $< 100$ Jm$^{-2}$s$^{-1/2}$K$^{-1}$ \citep{Hanus2018}. If the grain size plays a significant role on the asteroids, we would find the variation in spectral slope among different size groups. However, we did not find a significant difference between $d < 10$ km group and $d>10$ km group in visible spectral slope (Fig. \ref{fig:uvdist}). This might suggest that the grain size is not the primary cause to change the spectral slope on asteroid surfaces.

Polarimetry is also a powerful tool for estimating an asteroid’s surface condition. The linear polarization degree can be measured as a function of phase angle \citep[see][as a review]{Belskaya2015}. The linear polarization degree of asteroids reaches minimum value ($P_{\rm min}$) at a certain phase angle ($\alpha_{\rm min}$). The primitive bodies, such as D-class asteroids and the dark side of Iapetus, exhibit the smallest $|P_{\rm min}|$ \citep{Hasegawa2021}, while the asteroids with 0.7-$\mu$m band (Ch,G) have the largest $\alpha_{\rm min}$ and $|P_{\rm min}|$ \citep{Belskaya2017}. F types have values in-between these two end members, characterized by relatively small $\alpha_{\rm min}$ and $|P_{\rm min}|$ compared with other C-complex asteroids \citep{Belskaya2005, Belskaya2017, Gil-Hutton&Canada-Assandri2012, Hasegawa2021}. Asteroids classified as C and P types have similar inversion angles (the angle at which the linear polarization changes its sign) but different depths of negative polarization \citep{Belskaya2017}. It should be noted that asteroid spectral taxonomic classes defined by the reflectance in NUV-VIS space were well separated in polarization space \citep{Belskaya2017}. Although F and B types sometimes have similar negative visible slopes, they are well separated in polarization space. If the grain size plays a dominant role to determine the polarization feature, taxonomic classes should align in the same order of visible slope values. However, it is not the case. This suggests that the polarization features are related to the differences in composition rather than particle sizes or physical conditions. 

The interpretation of $P_{\rm min}$ and the depth of the negative branch have not been fully understood. The coherent back-scatter enhancement has been considered by several authors as the most plausible cause of negative polarization \citep{Shkuratov1985, Shkuratov1991, Shkuratov1994, Muinonen1990, Mishchenko1993}. More recent laboratory and theoretical studies showed a considerable role in the formation of negative polarization by single-particle scattering \citep{Ovcharenko2001, Shkuratov2004}. A few experimental studies suggested that adding small amounts of a bright material to a dark material might enhance the negative polarization branch, when bright particles have more prominent negative polarization than the dark absorbing particles \citep{Shkuratov2004, Belskaya2005}. The relatively small values of the parameters $\alpha_{\rm min}$ and $|P_{\rm min}|$ for D/F-type asteroids can be treated as a diagnostic of optical homogeneity and darkness of the regolith micro-structure on scales of the order of the wavelength \citep{Belskaya2005}, while large values of $\alpha_{\rm min}$ and $|P_{\rm min}|$ for Ch/G-type asteroids can be explained as inhomogeneity caused by bright inclusions such as Calcium-Aluminum-rich Inclusions and chondrules \citep{Devogele2017}. The peculiar polarimetric characteristics of the F type could also contribute to the NUV reflectance behavior. Thus, it is important to classify asteroids taking into account the polarimetric behavior in the future.

\subsection{F-type asteroids: Relation with (162173) Ryugu and (101955) Bennu}\label{sec:ryugu}
F types can only be classified using the NUV wavelengths; they are lately confused with B/C types in the visible wavelength range. The F-type asteroids have a peculiar distribution compared with other types of primitive asteroids (Fig. \ref{fig:semimajor_tax}): they concentrate in the IMB. Moreover, the Hayabusa2 spacecraft recently visited asteroid (162173) Ryugu, which is an F-type asteroid \citep{Tatsumi2022}. Ryugu was considered to originate from the IMB \citep{Campins2010} and more specifically from the Polana-Eulalia family \citep{Campins2013, Sugita2019, Tatsumi2021}. Thus, the Ryugu samples could be mineralogical representative of the F-type asteroids. From the analysis of the returned samples, Ryugu, which shows a strong and sharp OH absorption at 2.72 $\mu$m, has a composition similar to that of CI chondrites \citep{Yada2022, Pilorget2022, Yokoyama2022,Nakamura2022}. 

Based on the remote-sensing spectra, Ryugu was presumed to be a dehydrated carbonaceous chondrite due to its very flat and shallow absorption features in VIS-NIR \citep{Sugita2019, Kitazato2019, Tatsumi2021}. The case of Ryugu and CI chondrites led us to misinterpret the remote-sensing spectra due to the terrestrial contamination of CI chondrites and the space weathering of the Ryugu surface. This is a good lesson to learn and suggests that most of our charbonaceous chondrite meteorites might be contaminated and do not show exactly the right spectra to compare with asteroid surfaces. Nevertheless, CI chondrites are composed mainly of Mg-rich phyllosilicates and lack chondrules and CAIs \citep{Cloutis2011a}. Less abundance of Fe-righ pyllosilicates or presence of magnetite might result in the flat NUV reflectance in this case. Alternatively, the space weathering after the catastrophic disruption of the parent body might cause the shallower absorption features by phyllosilicates \citep{Noguchi2022}. These geochemical characteristics match the distant formation from the Sun. Moreover, the measurement of Cr isotopic heterogeneities suggests that CI chondrite Orgueil is formed farther from the Sun than Murchison (CM), Allende (CK) and Tagish Lake \citep{Fukai&Arakawa2021}. If F types were formed at a far distance from the Sun, they were implanted relatively recently into the current positions, mainly in the IMB. Another noticeable feature of F types is that they were formed relatively small in size. This could be the reason that they could be easily moved by disturbances due to the giant planets. The mechanism of displacement needs to be investigated further by dynamical calculations.

Furthermore, the asteroid (101955) Bennu visited and sampled by the OSIRIS-REx spacecraft is also classified as an F type \citep{Hergenrother2013}. Bennu could also originate from the Polana-Eulalia family in the IMB \citep{Campins2010, Bottke2015, Tatsumi&Popescu2021}. The returned samples from Bennu and Ryugu will add more geochemical information to our understanding of the F-type asteroids.

\section{Conclusions}

Phyllosilicates on primitive asteroids resulted from the reaction of aqueous alteration process in the early history of the Solar System. They are a key to constrain for the formation conditions of the primitive asteroids/planetesimals. The NUV absorption has been proposed to be a good proxy for phyllosilicates, but to date this had not been investigated well. For this reason, we explored the NUV region for dark primitive asteroids using two spectrophotometric surveys, SDSS and ECAS. Photometric surveys are more robust to investigate the sensitive NUV region than the spectroscopic observations from ground which can be affected greatly by the atmospheric conditions and the solar analogs. Using two surveys we can cover 9,168 asteroids with $\mathcal{H}<17.5$. 

First, we investigated the correlation between the NUV absorption and the OH-band (2.7 $\mu$m) absorption for asteroids and meteorites. We found a good correlation between the NUV and OH-band absorptions, which was originally discussed by \citet{Feierberg1985} and \citet{Hiroi1993}. We found that grain size may contribute the deeper band absorptions. Also, Fe-bearing phyllosilicates which also show the 0.7-$\mu$m band absorption may contribute the deeper NUV absorption. Based on these correlations, we confirmed that the NUV absorption is a good proxy for the phyllosilicate abundance on asteroids.

\citet{Tholen1984}'s taxonomic classification of asteroids has taken into account the NUV absorption. Following the Tholen taxonomy, we classified the asteroids and found their distribution in the main belt to the Cybele, Hilda, and Jupiter Trojan zones. We found that large asteroids with $d > 50$ km and small asteroids with $d< 10$ km have a completely different taxonomic distribution. Large asteroids show quite similar distribution through the main belt, while small asteroids exhibit different distributions at the inner main belt vs. the middle and outer main belt. The inner main belt region shows significantly small value of the NUV absorption and is dominated by F types. On the contrary, the major constituent of the middle and outer main belt are G types, which have the largest NUV absorption. The Cybele, Hilda, and Jupiter Trojan zones show distinctive distributions, as being dominated by the red members, P and D types. F types are found to be abundant also in the small members of the Cybele zone.

We found that the distribution of F types is unique: they concentrate in small members of inner main belt region. There is still not much constraint on the composition of F types. Recent sample-return missions, Hayabusa2 to (162173) Ryugu and OSIRIS-REx to (101955) Bennu, will also add fundamental chemical information on F types. So far the Ryugu samples are discussed to be compositionally similar to CI chondrites. Thus, the main constituent of F types may be CI-like Mg-rich phyllosilicates. Furthermore, the Gaia space telescope spectroscopically have observed numerous asteroids in this missing NUV wavelength. Gaia will open a new horizon to the compositional distribution using the NUV wavelength range.

\begin{acknowledgements}
The authors thank the anonymous reviewer for the constructive and elaborated comments and suggestions. ET, FTR, JdL, and JL acknowledge support from the Agencia Estatal de Investigaci\'{o}n del Ministerio de Ciencia e Innovaci\'{o}n 
(AEI-MCINN) under grant "Hydrated minerals and organic compounds in primitive asteroids" (PID2020-120464GB-I00/doi:\url{10.13039/501100011033}). JdL also acknowledges financial support from the Spanish Ministry of Science and Innovation (MICINN) through the Spanish State Research Agency, under Severo Ochoa Program 2020-2023 (CEX2019-000920-S). MP was supported by the grant of the Romanian National Authority for Scientific Research - UEFISCDI, project No. PN-III-P1-1.1-TE-2019-1504. SH was supported by the Hypervelocity Impact Facility (former name: The Space Plasma Laboratory), ISAS, JAXA.
\end{acknowledgements}

\bibliographystyle{aa} 
\bibliography{nuv2.bib} 
%

\end{document}